\documentclass[12pt,preprint]{aastex}

\usepackage{natbib}
\shorttitle{Magnetic Field of NGC 4214}
\shortauthors{Kepley et al.}

\newcommand{\ha}{\ensuremath{{\rm H}\alpha}}
\newcommand{\kms}{\ensuremath{\rm{km \, s^{-1}}}}

\newcommand{\uG}{\ensuremath{\rm{\mu  G}}}
\newcommand{\alfven}{Alfv{\'e}n}
\newcommand{\cms}{\ensuremath{{\rm cm \, s^{-1}}}}

\begin{document}

\title{The Magnetic Field of the Irregular Galaxy NGC 4214}

\author{Amanda A. Kepley\altaffilmark{1}, Ellen G. Zweibel, Eric M. Wilcots}
\affil{Washburn Observatory, University of Wisconsin, 475 North Charter Street, Madison, WI 53706}
\email{kepley@astro.wisc.edu,zweibel@astro.wisc.edu,ewilcots@astro.wisc.edu}

\altaffiltext{1}{Now at Department of Astronomy, University of
  Virginia, P.O. Box 400325, Charlottesville, VA 22904, USA; Visiting
  Research Associate at National Radio Astronomy Observatory, 520
  Edgemont Road, Charlottesville, VA 22903, USA.}

\author{Kelsey E. Johnson}
\affil{Department of Astronomy, University of Virginia, P.O. Box 400325,
Charlottesville, VA 22904, USA ; Adjunct at National Radio Astronomy Observatory, 520 Edgemont Road, Charlottesville, VA 22903, USA.} 
\email{kej7a@virginia.edu}

\author{Timothy Robishaw}
\affil{Sydney Institute for Astronomy, School of Physics, The University of Sydney, NSW 2006, Australia}
\email{tim.robishaw@sydney.edu.au}

\begin{abstract}
 
  We examine the magnetic field in NGC 4214, a nearby irregular
  galaxy, using multi-wavelength radio continuum polarization data
  from the Very Large Array. We find that the global radio continuum
  spectrum shows signs that free-free absorption and/or synchrotron
  losses may be important. The 3\,cm radio continuum morphology is
  similar to that of the \ha\, while the 20\,cm emission is more
  diffuse. We estimate that 50\% of the radio continuum emission in
  the center of the galaxy is thermal. Our estimate of the magnetic
  field strength is $30\pm 9.5$~\uG\ in the center and $10\pm3$~\uG\
  at the edges. We find that the hot gas, magnetic, and the
  gravitational pressures are all the same order of magnitude. Inside
  the central star forming regions, we find that the thermal and
  turbulent pressures of the HII regions dominate the pressure
  balance. We do not detect any significant polarization on size
  scales greater than 200~pc. We place an upper limit of 8~\uG\ on the
  uniform field strength in this galaxy. We suggest that the diffuse
  synchrotron region, seen to the north of the main body of emission
  at 20\,cm, is elongated due to a uniform magnetic field with a
  maximum field strength of 7.6~\uG. We find that, while the shear in
  NGC 4214 is comparable to that of the Milky Way, the supernova rate
  is half that of the Milky Way and suggest that the star formation
  episode in NGC 4214 needs additional time to build up enough
  turbulence to drive an $\alpha-\omega$ dynamo.

\end{abstract}


\section{Introduction}\label{ch3:sec:introduction}

Modern theories of galaxy formation postulate a bottom-up process
where smaller systems merge together to form successively larger
structures like the spiral galaxies we see today
\citep{ch3:2005Natur.435..629S}.  In the local universe, low-mass
galaxies are analogs to these high-redshift ``building blocks.'' By
studying nearby low mass galaxies, in particular low mass irregular
galaxies, we can gain detailed insight into the basic physical
processes at work in these systems, which are different than those in
spirals or ellipticals due to their smaller mass. Studying the
interstellar medium (ISM) of irregular galaxies is particularly
important because their ISM is strongly influenced by star formation
and interactions, which significantly affect the future evolution of
these galaxies.

There have been a number of surveys of low mass irregular galaxies
aimed at investigating the properties of the neutral and ionized gas
components of the ISM and the contributions to the ISM by the stellar
components (e.g.,
\citealt{ch3:hunter2004,ch3:hunter2006ir,ch3:hunter2006ubv,ch3:2007AAS...211.9506H,2010AJ....139..447H}).
Relatively little attention, however, has been paid to the magnetic
field component of the interstellar medium.

In general, the magnetic field energy is thought to be in
equipartition with the turbulent and cosmic ray energies
(\citealp{ch3:zh97,2001RvMP...73.1031F,2005ARA&A..43..337C}; although
see \citealp{ch3:cw1993} for a contrasting view) and is thus an
important part of the interstellar medium. In the few irregular
galaxies with measured magnetic fields
\citep{ch3:1986A&A...159...22H,ch3:1991A&A...252..475H,ch3:chyzy2000,ch3:chyzy2000erratum,ch3:chyzy2003,ch3:gaensler2005,ch3:mao2008,2010ApJ...712..536K,2011A&A...529A..94C},
the magnetic fields are comparable in strength to those in larger
spirals like the Milky Way
\citep{2001RvMP...73.1031F,2005LNP...664...89W} and M51
\citep{1987A&A...186...95B}. Therefore, in irregular galaxies, the
magnetic field may play a more important role in the dynamics of the
ISM than in spiral galaxies. Magnetic fields also channel gas flows
and distribute and accelerate cosmic rays
\citep{ch3:beck2005,2008ApJ...674..258E,2010ApJ...711...13E}. Irregular
galaxies have shallow potential wells that are easily disrupted and
potentially have a significant amount of their mass expelled from the
galaxy \citep{2000MNRAS.313..291F,2004ApJ...613..898T}. Magnetic
fields may either help the baryons escape or help to confine them to
the galaxy depending on the structure of the magnetic fields
\citep{ch2:1990ApJ...361L...5T,ch2:1993ApJ...409..663M,ch2:2005A&A...436..585D,2008ApJ...674..258E,2010ApJ...711...13E}. Finally,
the presence of significant magnetic fields in irregular galaxies
raises the question of how they were generated in the first place. The
main ingredients in generating and sustaining galactic magnetic fields
are generally thought to be large-scale shear and small-scale
turbulence. The well known $\alpha-\omega$ dynamo models are based on
these effects (although see \citealt{ch3:zh97},
\citealt{1999ARA&A..37...37K}, , and \citealt{2008RPPh...71d6901K} for
critiques of this model). In the $\alpha-\omega$ dynamo, differential
rotation ($\omega$) is used to stretch small-scale turbulence
($\alpha$) leading to an increase in magnetic field strength and the
presence of a large-scale magnetic field. It may be harder for
$\alpha-\omega$ dynamos to operate in irregular galaxies because they
rotate mostly as solid bodies and their ISM is more easily disrupted
because of their smaller potential wells. The strength and structure
of the magnetic fields in irregular galaxies provide stringent
constraints for dynamo models of these objects.

At present, only a handful of nearby galaxies have detailed magnetic
field measurements: NGC 1569 \citep{2010ApJ...712..536K}, the Large
and Small Magellanic Clouds
\citep{ch3:1986A&A...159...22H,ch3:1991A&A...252..475H,ch3:1993A&A...271..402K,ch3:gaensler2005,ch3:mao2008},
NGC 4449 \citep{ch3:chyzy2000}, IC 10 \citep{ch3:chyzy2003}, NGC 6822
\citep{ch3:chyzy2003}, and IC 1613 \citep{2011A&A...529A..94C}. To
better understand the role of magnetic fields in the interstellar
medium of irregular galaxies and how these fields are generated in
these systems, we have embarked on a program to obtain deep,
high-resolution radio continuum data on several irregular galaxies. In
this first paper in this series \citep{2010ApJ...712..536K}, we
presented our observations of the magnetic field of NGC 1569, a dwarf
irregular galaxy hosting one of the most extreme starbursts in the
nearby universe. The magnetic field in this galaxy is shaped by its
outflow and the field may be playing an important role in channeling
gas away from the disk.

Work by other groups have found that the LMC, SMC, and NGC 4449 have
large-scale ordered fields like NGC 1569. The LMC has a 1~\uG\
regular, axisymmetric spiral field, a random field of 4.1~\uG, and a
total field of 4.3~\uG\ \citep{ch3:gaensler2005}. A cosmic ray-driven
dynamo \citep{ch3:1992ApJ...401..137P,ch3:2004ApJ...605L..33H} has
been invoked to explain the presence of a regular field despite the
vigorous recent star formation in the LMC
\citep{ch3:gaensler2005}. The SMC has a random magnetic field strength
of 3~\uG\ and a unidirectional, regular field with a strength of
1.7~\uG\ aligned roughly along the Magellanic bridge, which suggests
an origin associated with the interaction between the LMC and SMC
\citep{ch3:mao2008}. In contrast to the weak, but regular fields in
the LMC and SMC, NGC 4449 has a strong uniform field of 8~\uG\ and a
total field of 14~\uG\ \citep{ch3:chyzy2000}. The uniform field shows
a fan-like structure in the inner regions of the galaxy and a spiral
structure in the outer regions. A fast dynamo model including velocity
shear around a bar, a wind from the central region of the galaxy, and
a random field produces a magnetic field configuration very similar to
that of NGC 4449 \citep{ch3:om2000}. In contrast to the magnetic
fields of NGC 1569, the LMC, the SMC, and NCG 4449, the uniform fields
of IC 10, NGC 6822, and IC 1613 (as traced by the polarization of
their radio continuum emission) are either weak or non-existent
\citep{ch3:chyzy2003}. IC 10 has a diffuse, random field with a
strength of 14~\uG\ in the center and 7~\uG\ on its outskirts. NGC
6822 has a random field with a maximum strength of 5~\uG\ and IC 1613
has a random magnetic field of 2.8~\uG. Both IC 10 and NGC 6822 have
small patches of uniform fields with strengths of 2--3~\uG\ on size
scales much smaller than the size of the galaxy. \citet{ch3:chyzy2003}
suggest that the fields in these galaxies are the result of a
fluctuating dynamo \citep{ch3:1998MNRAS.294..718S}. IC 1613 shows no
sign of uniform field; any polarized emission can be attributed to
background sources \citep{2011A&A...529A..94C}.

The line of sight fields determined by \citet{ch3:gaensler2005} and
\citet{ch3:mao2008} for the LMC and SMC are based on the rotation
measures of background sources, while the results for NGC 4449, IC 10,
NGC 6822, IC 1613, and NGC 1569 are based on observations of the radio
continuum polarization. There are significant differences between the
two methods. The radio continuum polarization observations trace the
magnetic field with cosmic ray electrons, while the rotation measure
observations trace the field with thermal electrons. In addition, only
the rotation measure observations can tell a regular (i.e., coherent)
field (all one direction) from a uniform field (one orientation, but
the direction uncertain by 180$^\circ$). Unfortunately, current
technology limits this method to nearby galaxies (e.g., LMC, SMC, M31)
with very large angular sizes, although this situation will improve
with the development of the Square Kilometer Array
\citep{2004NewAR..48.1289B}.

In this paper, we focus on NGC 4214, which is classified as a mixed,
S-shaped, Magellanic irregular galaxy by
\citet{ch3:1995yCat.7155....0D}. NGC 4214 is a well-studied galaxy
whose rest-frame UV spectrum strongly resembles that of star-forming
galaxies at $z=3$ \citep{ch3:1996ApJ...462L..17S}. Observations of its
stars and ionized gas
\citep{ch3:1998A&A...329..409M,ch3:2000AJ....120.3007M,ch3:2002AJ....124..811D,ch3:2007AJ....133..917U,ch3:2007AJ....133..932U}
show that there are two active regions of star formation in the
galaxy: NGC 4214-I (also referred to as the northwest complex) and NGC
4214-II (also referred to as the southeast complex). NGC 4214-I is the
older complex with an age of 3 to 4~Myr and consists of an \ha\ shell
surrounding a super-star cluster \citep{ch3:2000AJ....120.3007M}.  The
\ha\ kinematics of NGC 4214-I show some indications of expanding
shells
\citep{ch3:martin98,ch3:1999A&A...343...64M,ch3:2001ApJ...555..758W}. NGC
4214-II is slightly younger with an age of 2.5 to 3.0~Myr
\citep{ch3:2000AJ....120.3007M}. The \ha\ in this region is coincident
with the star cluster, showing that the cluster has not had sufficient
time to clear out a large \ha\ cavity
\citep{ch3:2000AJ....120.3007M}. \citet{ch3:schwartz04} detect an
outflow of cold gas in this region with a velocity of 23~$\rm{km \
s}^{-1}$, which is associated with a small \ha\ bubble. X-ray
observations with Chandra and XMM-Newton show that the hot gas in this
galaxy is centered on the star forming regions
\citep{ch3:2004MNRAS.348..406H,ch3:2005MNRAS.358.1423O,ch3:ott05II}. The
star forming regions of NGC 4214 and their associated ionized and hot
gas are embedded in a disk of much older stars
\citep{ch3:1997ApJ...481..735F,ch3:2002AJ....124..811D}, which in turn
is embedded in a neutral hydrogen envelope which extends to 1.4 times
the Holmberg radius (5.3\arcmin)\footnote{The Holmberg radius is the
radius where the surface brightness of a galaxy falls to 26.5
mag~$\rm{arcsec}^{-2}$.} of this galaxy
\citep{ch3:1979MNRAS.188..765A,ch3:mcintyre97}.

The distance to NGC 4214 is controversial, with distances used in the
literature ranging from 2~Mpc \citep{ch3:1999ASPC..192...85H} to 7~Mpc
\citep{ch3:1979MNRAS.188..765A}. We adopt a distance of 2.94~Mpc in
this paper, which is the distance for the galaxy determined from
Hubble Space Telescope imaging of the tip of the red giant branch by
\citet{ch3:2002AJ....123.1307M}. This distance agrees with the tip of
the red giant branch distances determined independently by
\citet[2.70~Mpc]{ch3:2002AJ....124..811D} and
\citet[3.04~Mpc]{2009ApJS..183...67D} and with the distance determined
using the planetary nebula luminosity function by
\citet[3.19~Mpc]{2010Ap&SS.tmp..148D}. It is also in reasonable
agreement with distances determined by its radial velocity and models
of the Hubble flow determined by NED\footnote{The NASA/IPAC
  Extragalactic Database (NED) is operated by the Jet Propulsion
  Laboratory, California Institute of Technology, under contract with
  the National Aeronautics and Space Administration.} (3.75 to
7.53~Mpc). At the adopted distance, 1\arcsec\ is 14.2~pc.

The observations presented in this paper represent the most sensitive
observations of the radio continuum emission of NGC 4214 to
date. Previous observations of the radio continuum of NGC 4214 include
\citet{ch3:1979MNRAS.188..765A}, \citet{ch3:2000AJ....120..244B}, and
\citet{ch3:2000AJ....120.3007M}. Although the resolution of our
observations is lower than that of \citet{ch3:2000AJ....120..244B} and
\citet{ch3:2000AJ....120.3007M}, we integrated for substantially
longer (effective integration times ranging from 5 hours to 15 hours
rather than 15 minutes). Our choice of array configuration (see
Section~\ref{ch3:sec:data}) allows us good sensitivity to both small
scale features and large-scale diffuse emission in NGC 4214. Our
observations trace the thermal emission produced by the ionizing
radiation of young stars and the synchrotron emission produced by the
detonation of old massive stars, which nicely complements the new data
on the stellar content of NGC 4214 obtained by the Wide Field Camera 3
on the Hubble Space Telescope \citep[e.g.,][]{2010Ap&SS.tmp..148D} and
the new data on the molecular components of the ISM obtained using
Herschel \citep[e.g.,][]{2010A&A...518L..57C}.

We detail our data calibration and imaging process in
Section~\ref{ch3:sec:data}. In Section \ref{ch3:sec:radio-cont-emiss},
we present our observations in
detail. Section~\ref{ch3:sec:discussion} is devoted to our results
including a comparison with multi-wavelength data
(Section~\ref{sec:comp-with-emiss}), estimates of the thermal fraction
of the radio continuum emission (Section~\ref{sec:fract-therm-emiss}),
an estimation of the magnetic field strength in NGC 4214
(Section~\ref{ch3:sec:magn-field-strength}), an analysis of the
importance of the magnetic field in the ISM of NGC 4214
(Section~\ref{sec:import-magn-field}), an investigation of the
magnetic field structure of NGC 4214
(Section~\ref{sec:struct-magn-field}), estimates of cosmic ray
lifetimes in NGC 4214 (Section~\ref{sec:cosic-ray-diffusion}), and a
discussion of possible mechanisms for generating the field
(Section~\ref{sec:gener-large-scale}). Finally, we summarize our
conclusions in Section~\ref{ch3:sec:conclusions}.

\section{Data}\label{ch3:sec:data}

We have obtained radio continuum polarization observations of NGC 4214
at 20, 6, and 3\,cm using the National Radio Astronomy Observatory
(NRAO) Very Large Array (VLA).\footnote{The National Radio Astronomy
  Observatory is a facility of the National Science Foundation
  operated under cooperative agreement by Associated Universities,
  Inc.} Our observations (proposal IDs: AK606, AK616; PI: Kepley) were
taken between 2005 August and 2005 December. See
Table~\ref{tab:vla_obs_summary} for a summary of these
observations. We used two different configurations of the VLA to as
closely as possibly match the synthesized beams at our three different
observing frequencies. The correlator was configured for two
intermediate frequencies (IFs) each with 50~MHz bandwidth and full
polarization products (RR, LL, RL, LR). For all our observations, we
used 1331+305 (3C286) to calibrate the flux density scale and absolute
polarization angle and the relevant secondary calibrator to calibrate
the amplitudes and phases as well as the instrumental polarization.

For our 20\,cm observations, we observed our secondary calibrator
(1227+365) for four minutes for every thirty-five minutes on
source. The primary calibrator (1331+305) was observed for ten minutes
near the beginning of the observations and ten minutes at the end of
the observations. For our 6\,cm observations, we observed each
pointing of a four pointing mosaic for eight minutes and observed the
secondary calibrator for four minutes once every cycle through the
mosaic. The primary calibrator was observed in a fashion similar to
that of our 20\,cm observations. For our 3\,cm observations, we
observed each pointing of a sixteen pointing mosaic for six
minutes. Every half cycle through the mosaic we observed our secondary
calibrator (1146+399) for six minutes. The primary calibrator
observations were the same as in our 20\,cm and 6\,cm observations.

We calibrated the data in AIPS. Our data reduction procedures followed
the standard procedures to calibrate VLA data outlined in the AIPS
Cookbook \citep{ch3:aipscookbook} with some additional effort to
properly calibrate the polarization. We briefly sketch our data
reduction procedures below.

After the initial flagging of the data, we used the primary calibrator
(with the appropriate {\em u-v} limits) to set the primary flux
density scale.  With our observational setup, 3C286 (1331+305) is
well-approximated as a point source. We then calibrated the amplitudes
and phases of the primary and secondary calibrators.  The flux density
value for the secondary calibrator was bootstrapped from the flux
density value obtained for the primary calibrator. We applied the
calibrated amplitudes and phases to their respective calibrators and
the calibrated amplitudes and phases from the secondary calibrator to
the source. To calibrate the instrumental polarization response, we
used our observations of the secondary calibrator. The secondary
calibrator was observed at least five times over at least 100 degrees
of parallactic angle, which allows us to use the rotation of the feed
with respect to the sky to separate the instrumental polarization
(which does not rotate as the parallactic angle changes) from the
astronomical polarization (which does rotate as the parallactic angle
changes). The calibration for the instrumental polarization was then
applied to the data. The primary calibrator was used to determine the
absolute polarization angle and this final correction applied to the
data.

After calibrating in AIPS, we exported the I, Q, and U data as FITS
files and imaged the data in Miriad \citep{ch3:1995ASPC...77..433S} to
take advantage its ability to jointly deconvolve mosaics. For the
3\,cm and 6\,cm data, we used the {\em mosaic} option on the {\em
invert} task to combine the different pointings of the mosiac. The
data was weighted with a robust value of 0. For the 3\,cm data, we
tapered the data in the {\em u-v} plane to match the resolution of the
data at 20\,cm and 6\,cm. We then cleaned the mosaic using the task
{\em mossdi}. Since the 20\,cm data was only a single pointing, we
used {\em invert} without the {\em mosaic} option to image the data
and the task {\em clean} to clean the resulting image. In the 20\,cm
data, there was a bright source to the southwest of the galaxy that we
were not able to completely model during cleaning. The incomplete
removal of the source has the side-effect of increasing the noise in
the 20\,cm Stokes I image to approximately 4 times the theoretical
noise at 20\,cm and twice the noise in the other images. The Stokes Q
and U images were not affected as severely and their noise levels are
much closer to the theoretical
noise. Table~\ref{tab:final_image_summary} summarizes the properties
of our final images.

\section{Properties of the Radio Continuum Emission} \label{ch3:sec:radio-cont-emiss}

\subsection{Total Intensity} \label{sec:total-intensity}

\begin{figure}
\centering
\includegraphics[scale=1.0]{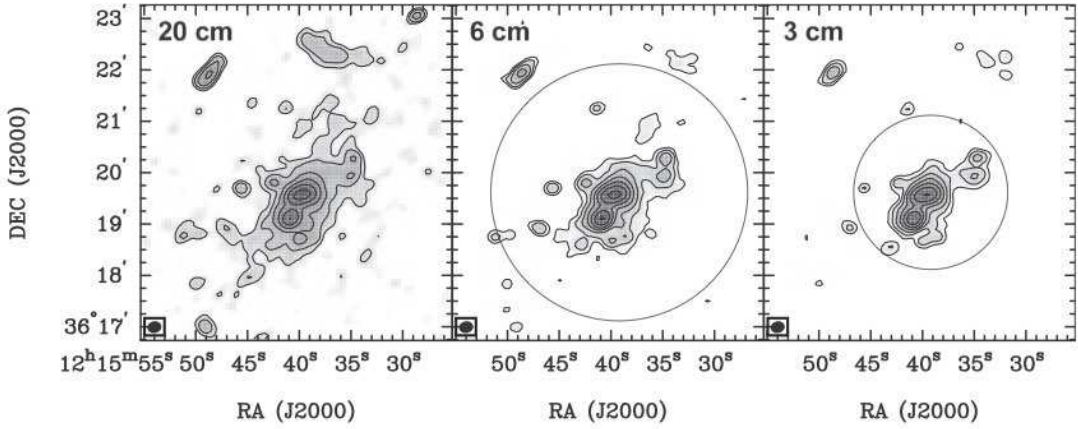}
\caption[NGC 4214 Radio Continuum Emission]{ Maps of the total
  intensity radio continuum emission of NGC 4214 at 20\,cm (left
  panel), 6\,cm (middle panel), and 3\,cm (right panel). The images
  are all on the same logarithmic scale stretching from 72.3 $\mu
  \rm{Jy}\ \rm{beam}^{-1}$ to 20 $\rm{mJy \ beam}^{-1}$. The contours
  are 3, 6, 12, 24, 48, 96, 192, and 384 times the 1$\sigma$ noise
  level given in Table~\ref{tab:final_image_summary}. The beam is
  boxed in the lower left corner of each panel and is identical for
  all three images. The diameter of the circles in the middle and
  right panels show the largest angular scale imaged by the VLA at
  6\,cm and 3\,cm, respectively. The largest angular scale imaged by
  the VLA at 20\,cm is larger than the size of the region shown.}
\label{ch3:fig:total_intensity}
\end{figure}

Figure~\ref{ch3:fig:total_intensity} shows the distribution of the
radio continuum emission in NGC 4214. The emission is most extended at
20\,cm despite the higher noise levels in the 20\,cm image. At 3\,cm,
the emission is largely confined to two regions near the center of the
galaxy. The circles in the middle and right panels of
Figure~\ref{ch3:fig:total_intensity} show the largest angular scale
imaged by the VLA at 6\,cm and 3\,cm. The largest angular scale imaged
by the VLA at 20\,cm is larger than the region shown in
Figure~\ref{ch3:fig:total_intensity}.

\begin{figure}
\centering
\includegraphics[scale=0.8]{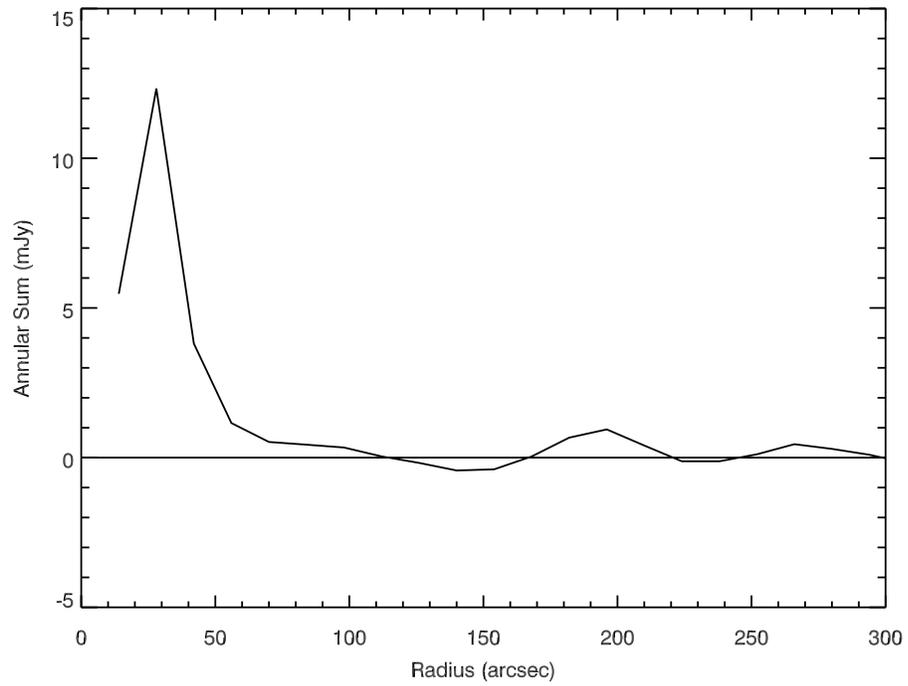}
\caption{3\,cm radio continuum emission per 14\arcsec\ annulus about
the center of the galaxy as a function of radius. The presence of a
large negative bowl of emission, which is not seen here, would
indicate that significant flux is resolved out.}
\label{fig:ellint}
\end{figure}

\begin{figure}
\centering
\includegraphics[scale=0.8]{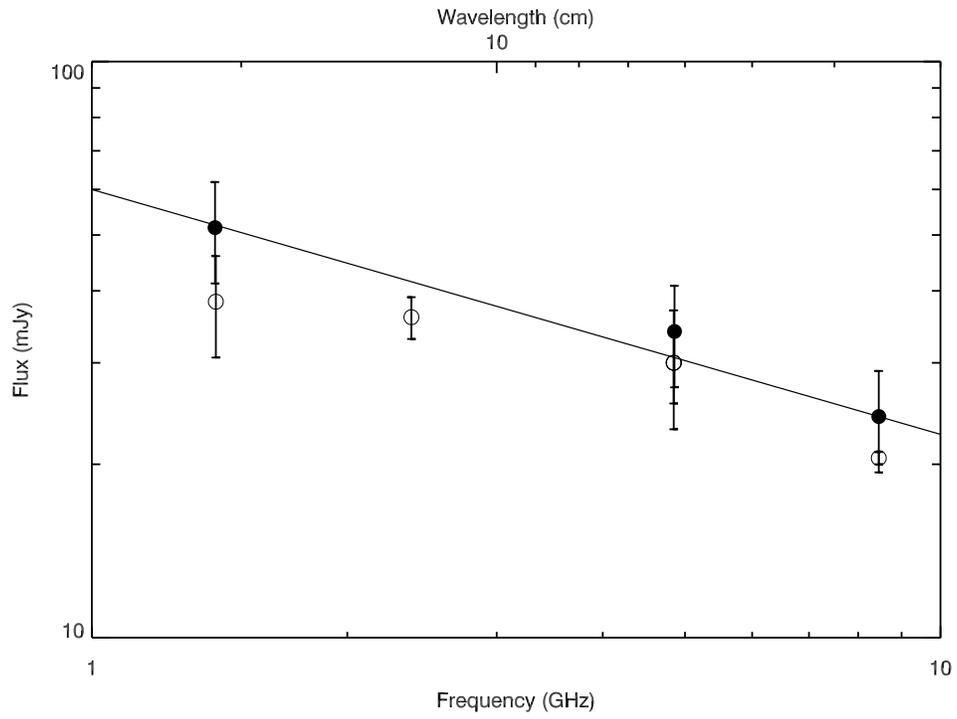}
\caption[Radio Continuum Spectrum of NGC 4214]{ Radio continuum
  spectrum of NGC 4214. The open points are previous measurements and
  the solid points are our measurements from this paper. The line is
  the best power law fit to our measurements; see the discussion in
  \S~\ref{sec:total-intensity}.}
\label{ch3:fig:radio_cont_spectrum}
\end{figure}

We recover most of the radio continuum flux from the galaxy. The first
evidence in favor of this conclusion is that the extent of NGC 4214 at
both 6\,cm and 3\,cm is much smaller than the largest angular scale
imaged at those wavelengths. The second line of evidence for this
conclusion is shown in Figure~\ref{fig:ellint}. In this Figure, we
plot the sum of the 3\,cm radio continuum emission in a particular
annulus as a function of radius from the center of the galaxy. The
presence of a large negative bowl surrounding the radio continuum
emission would indicate that large-scale emission is being resolved
out. This effect would be largest for the 3\,cm data because it will
resolve out emission on smaller scales than the 6\,cm or 20\,cm
data. There is no significant negative bowl surrounding our 3\,cm
radio continuum emission. The dip at 150\arcsec\ is well beyond the
main body of radio continuum emission and is the same magnitude as the
variations at larger radii. Finally, we have plotted the total flux
from our data along with measurements of the total flux from the
literature in Figure~\ref{ch3:fig:radio_cont_spectrum}. We assume that
the errors on our total fluxes and those of measurements from the
literature without quoted errors are 20\%. This error reflects the
fundamental uncertainty in the calibration of radio
fluxes. Table~\ref{tab:radio_cont_spectrum} lists the values and their
sources. Our measured radio continuum fluxes are larger than the
previously measured radio continuum fluxes. Therefore, we are not
resolving out significant emission on scales larger than our largest
angular scale. Our measured fluxes are larger than the fluxes from
previous observations because all of the previous observations were
shallow survey observations and these observations are able to detect
more faint diffuse emission.

Given the systematic offset between the archival measurements and our
data, the heterogeneity of the archival measurements, and the limited
number of measurements from our observations, we only fit a simple
model consisting of a single power law to our radio continuum spectrum
for NGC 4214. We find that the data can be fit by a single power law
with a spectral index\footnote{We define the spectral index, $\alpha$,
as $S_\nu \propto \nu^\alpha$.} equal to $-0.43 \pm 0.06$. This
spectral index shows that the spectrum includes a mix of thermal and
non-thermal emission. We see some hint that the spectrum is convex,
which would suggest that models including free-free absorption of
synchrotron and/or free-free emission or including synchrotron losses
might be appropriate \citep{1993ApJ...410..626D}. The electron
densities measured in NGC 4214 by \citet{ch3:1996ApJ...471..211K} are
similar to the electron densities determined by the modeling of the
radio continuum spectra of other similar galaxies including free-free
absorption of synchrotron emission by
\citet{1993ApJ...410..626D}. However, our limited data set does not
justify the use of more complex models. Additional data points at
higher and lower frequencies determined using data that matches the
spatial frequency range and sensitivity of our data are necessary to
model these more complex physical scenarios.

There are two isolated regions of radio continuum emission located to
the north and northeast of the main emission region. The first is a
region of diffuse 20\,cm emission to the north of the galaxy at
($12^{\rm h}15^{\rm m}38^{\rm s}, 36^\circ22\arcmin30\arcsec$),
referred to hereafter as the diffuse synchrotron region. This region
is likely to be part of NGC 4214 based on the presence of the optical
and \ha\ emission in the same location (see \S
\ref{sec:comp-with-emiss} for further discussion). The bright radio
continuum source with a steep radio continuum spectrum (see
\S~\ref{sec:spectral-indices}) at ($12^{\rm h}15^{\rm m}48.9^{\rm s},
36^\circ21\arcmin54.10\arcsec$) is resolved into two sources by the
Faint Images of the Radio Sky at Twenty centimeters (FIRST) survey
\citep{1995ApJ...450..559B}. \citet{2002ApJS..143....1M} is able to
identify an optical counterpart for this source on the red POSS-I
plates; the optical morphology is classified as blended (two local
maxima with a single set of connected above-threshold
pixels). Unfortunately, there are no determinations of the source
redshift. The southernmost of the two FIRST sources is coincident with
x-ray point source 20 from \citet{ch3:2004MNRAS.348..406H}. The x-ray
emission from this source is extremely hard ($(H-M)/(H+M) = 0.698$,
$(M-S)/(M+S) = 0.231$). Based on the steep spectra index in the radio,
the blended source identification of \citet{2002ApJS..143....1M}, and
the hard x-ray colors from the associated x-ray source, we conclude
that this source is a background AGN.

\subsection{Spectral Indices} \label{sec:spectral-indices}

\begin{figure}
\centering
\includegraphics[scale=1.0]{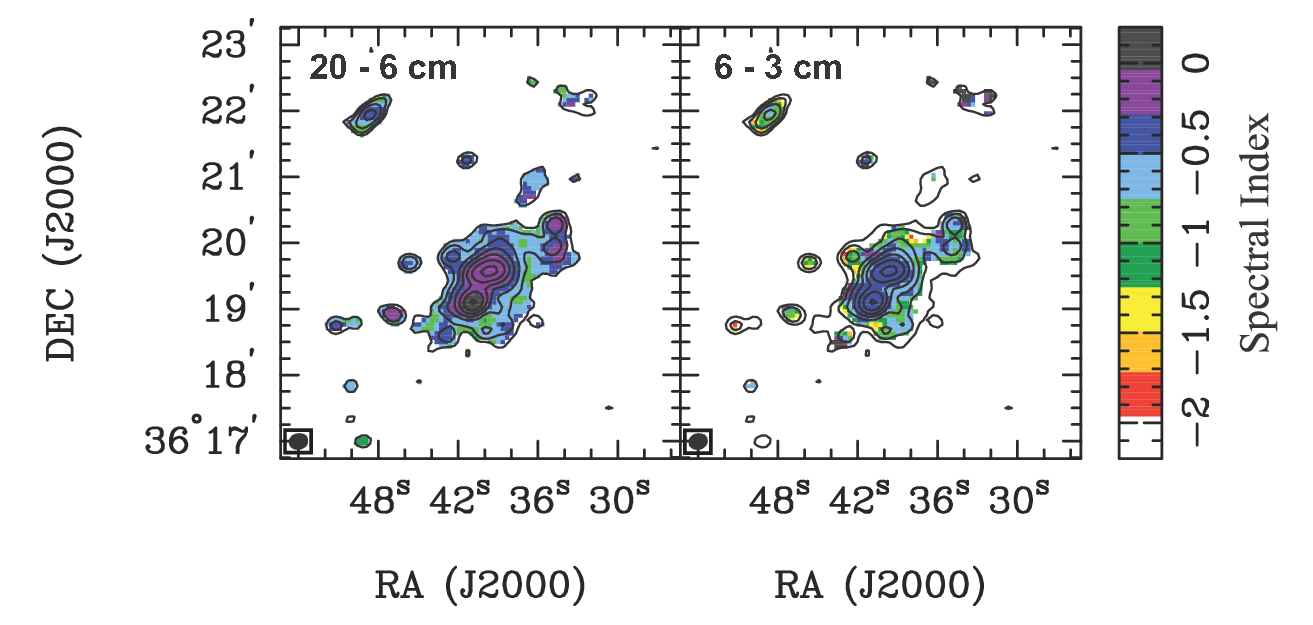}
\caption[Spectral Index maps for NGC 4214]{ Spectral index between the
  20\,cm and 6\,cm data (left panel) and between the 6\,cm and 3\,cm
  data (right panel). Pixels below 3$\sigma$ in the input images were
  masked before calculation of the spectral indices. The contours show
  the total intensity radio continuum emission at 6\,cm. The contours
  are 3, 6, 12, 24, 48, 96, 192, and 384 times the 1$\sigma$ noise
  level given in Table~\ref{tab:final_image_summary}. The beam is
  boxed in the lower left corner of each panel.}
\label{ch3:fig:spectral_index}
\end{figure}

The spectral indices between the 20\,cm and the 6\,cm data and the
6\,cm and the 3\,cm data are shown in
Figure~\ref{ch3:fig:spectral_index}. Regions in the input images below
3$\sigma$ were masked prior to calculating the spectral index. Thermal
emission has a spectral index of -0.1 and synchrotron emission has a
spectral index of $\sim -0.7$, although free-free absorption and
synchrotron losses can modify the observed spectrum. The two brightest
emission peaks are thermally dominated with the younger (2.5 to 3.0
Myr; \citealp{ch3:2000AJ....120.3007M}), southern peak (NGC 4214-II)
having a flatter spectral index than the older (3.0 to 4.0~Myr;
\citealp{ch3:2000AJ....120.3007M}), northern peak (NGC 4214-I). Flat
radio continuum spectra like those seen in NGC 4214-I and NGC 4214-II
are characteristic of dwarf starburst galaxies
\citep{1984A&A...141..241K,1993ApJ...410..626D}.  The two fainter
sources northwest of the galaxy (at 12$^{\rm h}$15$^{\rm m}$34$^{\rm
s}$, 36$\arcdeg$20$\arcmin$) are also thermally dominated, but with a
less steep spectrum than NGC 4214-I and NGC 4214-II. The regions
between NGC 4214-I and -II and the peaks to the northwest have steeper
spectra showing that synchrotron emission plays a more important role
in these regions.

\subsection{Polarized Emission} \label{sec:polarized-emission}

Since we were interested in investigating the structure of the
magnetic field, we determined the polarized intensity and the
polarization angle from the Stokes Q and U images. The derived
polarized intensities were corrected for the bias in the estimated
polarized intensity \citep{ch2:simmons1985,ch2:2006PASP..118.1340V}.

\begin{figure}
\centering
\includegraphics[scale=1.0]{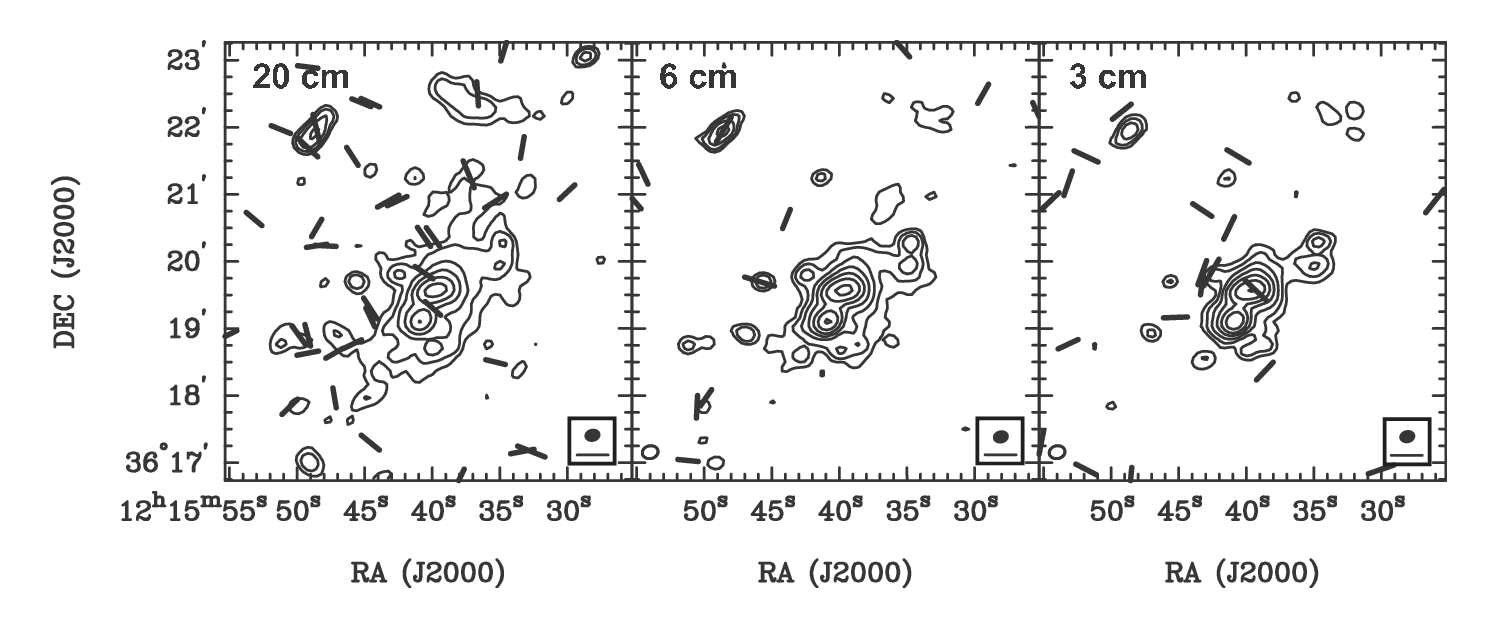}
\caption[NGC 4214 polarized intensity]{ Polarized intensity vectors
  (E-vectors) overlaid on contours showing the total intensity radio
  continuum emission of NGC 4214 at 20\,cm (left panel), 6\,cm (middle
  panel), and 3\,cm (right panel). The polarized intensities have been
  bias corrected and only vectors with a signal to noise greater than
  3 are shown.  The contours are 3, 6, 12, 24, 48, 96, 192, and 384
  times the 1$\sigma$ noise level given in
  Table~\ref{tab:final_image_summary}. The beam and the polarization
  vector scale bar are boxed in the right hand corner of each
  panel. The beam is identical for all three images. The length of the
  polarization vector scale bar corresponds to a polarized intensity
  of $5.5 \times 10^{-5} \ \rm{Jy \ beam}^{-1}$. }
\label{ch3:fig:polarized_intensity}
\end{figure}

The percent polarization gives the strength of the uniform field
relative to that of the total field, while the polarization angle
rotated by $\sim 90^\circ$ gives the orientation of the magnetic
field. The polarized emission of NGC 4214 is shown in
Figure~\ref{ch3:fig:polarized_intensity}. There is little polarized
emission associated with NGC 4214. This lack of polarization could be
due to several instrumental/imaging effects or due to depolarization
effects internal to the galaxy. The possible instrumental imaging
effects are: 1) depolarization due to mosaicing, 2) the polarized
emission is stronger at lower resolutions, or 3) bandwidth
depolarization due to combining two IFs. To check that mosaicing the
observations has not depolarized the final image or produced spurious
polarization signatures, we imaged each pointing separately at 6\,cm
and 3\,cm. The images of each pointing at both frequencies were
largely depolarized, so mosaicing was not the cause of the
depolarization seen in Figure~\ref{ch3:fig:polarized_intensity}.  To
see what effect decreasing the resolution of the images had we made
images with 1.5 times and 2.0 times the resolution of the VLA images
and an image with the resolution of a 100-m single dish telescope
(e.g., the Effelsberg 100-m telescope or the Green Bank
Telescope). Decreasing the resolution did not increase the amount of
visible polarization, so the resolution of these images was not the
cause of the depolarization. To see if bandwidth depolarization is the
cause of the depolarization, we imaged each IF separately and compared
their polarized emission and determined that there was no significant
depolarization associated with combining the two IFs for
imaging. Based on these experiments it does not seem likely that
instrumental/imaging effects are the cause of the depolarization.

The lack of polarization in Figure~\ref{ch3:fig:polarized_intensity}
could also be due to the properties of the galaxy itself including
depolarization effects internal to the galaxy or insufficient
sensitivity. Depolarization effects internal to the galaxy, e.g.,
\citet{ch3:1998MNRAS.299..189S}, decrease as ones goes to shorter
wavelengths. Therefore, the image at 3\,cm should be largely free of
depolarization effects while still having appreciable synchrotron
emission (see Sections~\ref{sec:total-intensity} and
\ref{ch3:sec:magn-field-strength}). The 3\,cm polarization image does
not show appreciably more polarization than the 20\,cm polarization
image, so internal depolarization effects do not seem to be leading to
depolarization at 20\,cm and 6\,cm. In
Section~\ref{sec:uniform-field-using}, we use our polarization data to
constrain the uniform magnetic field strength.

\subsection{Comparison with Previous Observations}\label{sec:comp-with-prev}

Our data represent the first observations of the polarized radio
continuum emission from NGC 4214. There are, however, three previous
studies of the radio continuum emission of NGC 4214:
\citet{ch3:1979MNRAS.188..765A}, \citet{ch3:2000AJ....120..244B}, and
\citet{ch3:2000AJ....120.3007M}.

The earliest interferometric observations of the radio continuum
emission from NGC 4214 were done at 20\,cm by
\citet{ch3:1979MNRAS.188..765A}. Despite their much larger beam ($58
\arcsec$ by $96 \arcsec$) and higher noise levels, the image derived
from these observations has the same overall shape as our higher
resolution and better sensitivity image. The emission to the southeast
at 12$^{\rm h}$15$^{\rm m}$34$^{\rm s}$, 36$\arcdeg$20$\arcmin$ that
they suggest is a background object is actually clearly part of the
main radio continuum distribution in our higher resolution
observations and is associated with faint \ha\ and optical emission.

\citet{ch3:2000AJ....120..244B} observed NGC 4214 with the VLA at
20\,cm, 6\,cm, 3.6\,cm, and 2\,cm with a resolution of $\sim 2
\arcsec$. By optimizing their observations for higher resolution,
however, they resolve out most of the extended structure seen in
Figure~\ref{ch3:fig:total_intensity}. \citet{ch3:2000AJ....120..244B},
however, are able to resolve NGC 4214-I and NGC 4214-II into several
point sources with hints of more extended emission, particularly in
NGC 4214-I. They find that the spectral indices that they derive
between 6\,cm and 2\,cm are rising, i.e., positive, and attribute this
to the free-free emission becoming optically thick. They estimate that
the turnover frequency for NGC 4214 is closer to 15~GHz than to
5~GHz. We do not identify any rising spectrum sources between 6 and
3\,cm. The beam of our lower resolution observations includes both
optically thick sources and extended emission so we would not expect
to be able to identify individual optically thick sources.

The 20\,cm and 6\,cm radio continuum observations in
\citet{ch3:2000AJ....120.3007M} are lower resolution ($\sim 4\arcsec$)
than those in \citet{ch3:2000AJ....120..244B}, but still higher
resolution than our observations. Again, the higher resolution of
these observations compared to ours allows them to resolve NGC 4214-I
and NGC 4214-II into several point sources with hints of extended
emission, but resolves out much of the diffuse
emission. \citet{ch3:2000AJ....120.3007M} were able to identify one of
the points sources in NGC 4214-I as a supernova remnant based on its
spectral index and the properties of the associated optical emission,
showing that supernovae have had enough time to produce appreciable
synchrotron radiation in NGC 4214.

\section{Discussion}\label{ch3:sec:discussion}

\subsection{Comparison of the Radio Continuum Emission with Emission at Other Wavelengths} \label{sec:comp-with-emiss}

\begin{figure}
\centering
\includegraphics[scale=0.6]{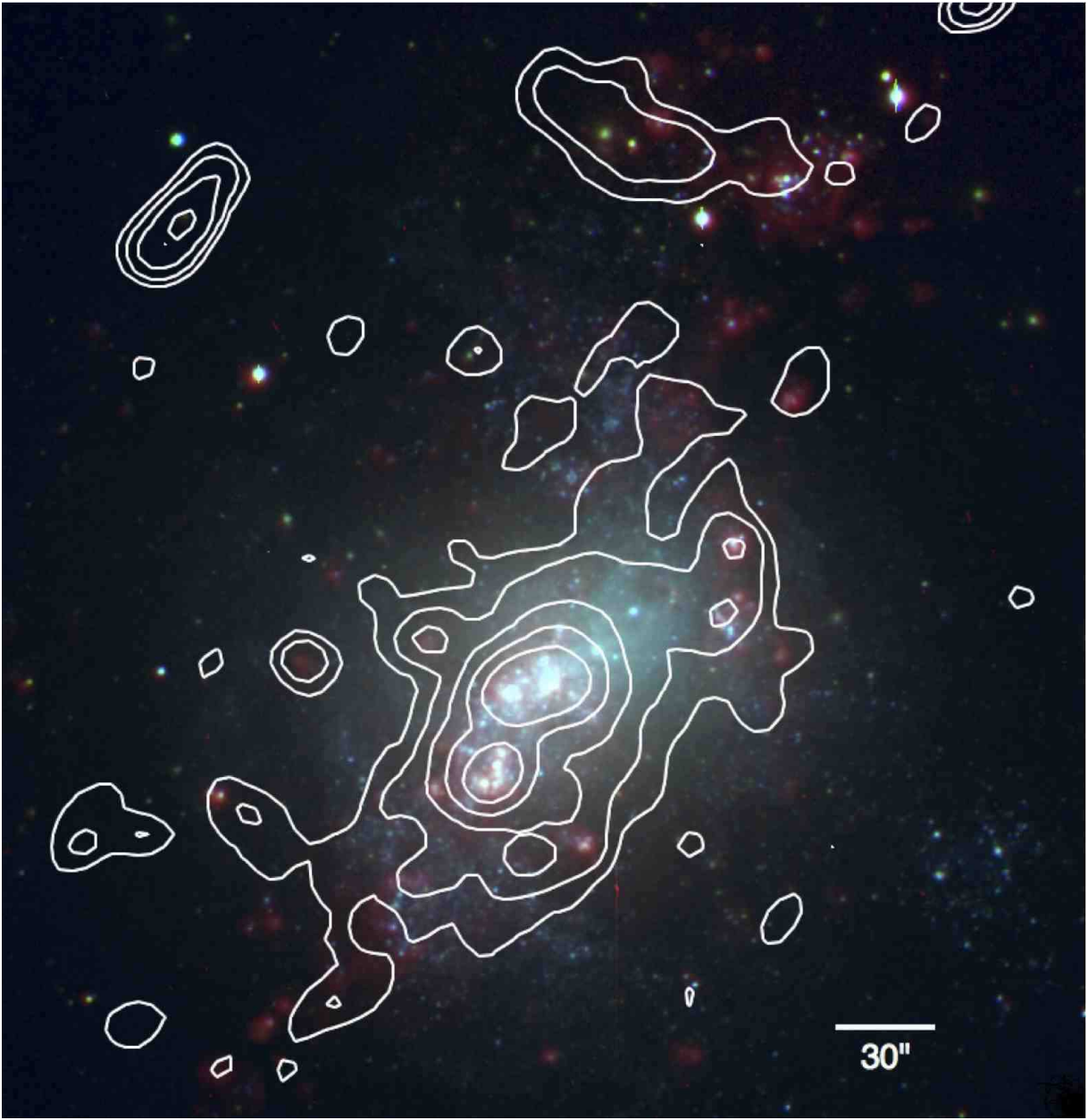}
\caption{20\,cm total intensity radio continuum emission contours
  overlaid on a three color optical image of NGC 4214. The optical
  images are courtesy D.\ Hunter and red represents \ha, blue the U
  band optical, and green V band optical. The contours are 3, 6, 12,
  24, 48, 96, 192, and 384 times the 1$\sigma$ noise level given in
  Table~\ref{tab:final_image_summary}. }
\label{fig:opt_lband}
\end{figure}

\begin{figure}
\centering
\includegraphics[scale=1.0]{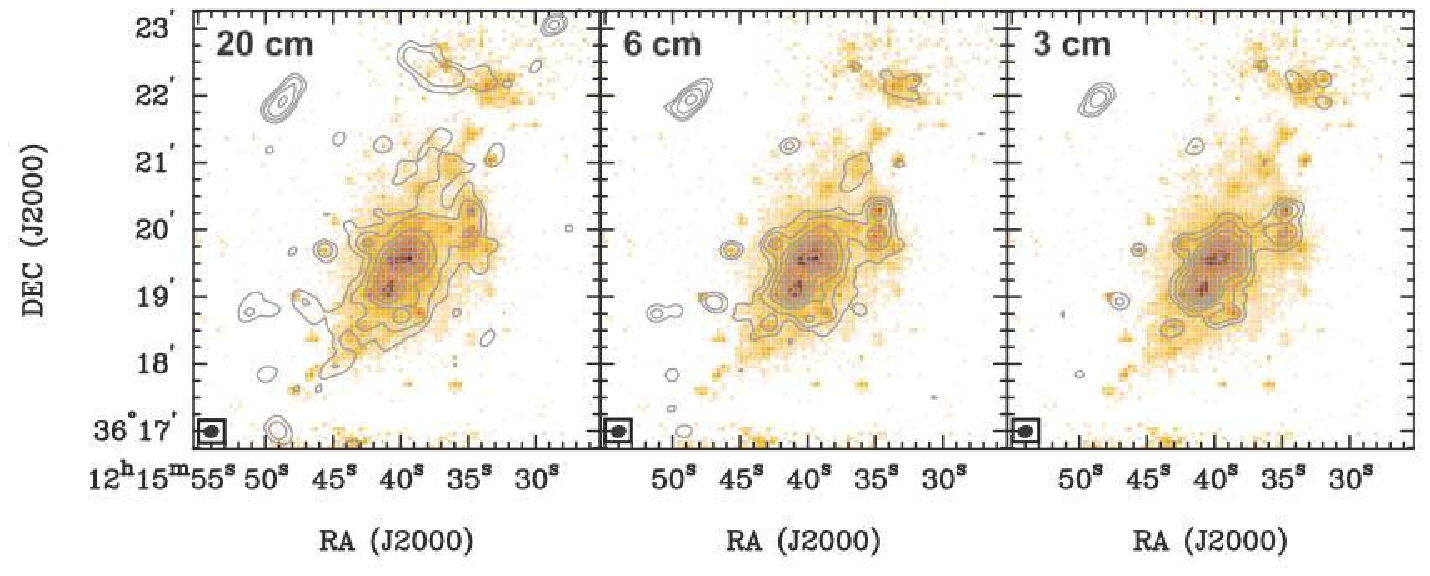}
\caption[Comparison of \ha\ and radio continuum emission of NGC 4214]{
  Contours of total intensity radio continuum emission at 20\,cm (left
  panel), 6\,cm (middle panel), and 3\,cm (right panel) overlaid on an
  \ha\ image of NGC 4214 kindly provided by Deidre Hunter
  \citep{ch3:hunter2004}. The contours are 3, 6, 12, 24, 48, 96, 192,
  and 384 times the 1$\sigma$ noise level given in
  Table~\ref{tab:final_image_summary}. The beam is boxed in the lower
  left corner of each panel. }
\label{ch3:fig:halpha}
\end{figure}

\begin{figure}
\centering
\includegraphics[scale=1.0]{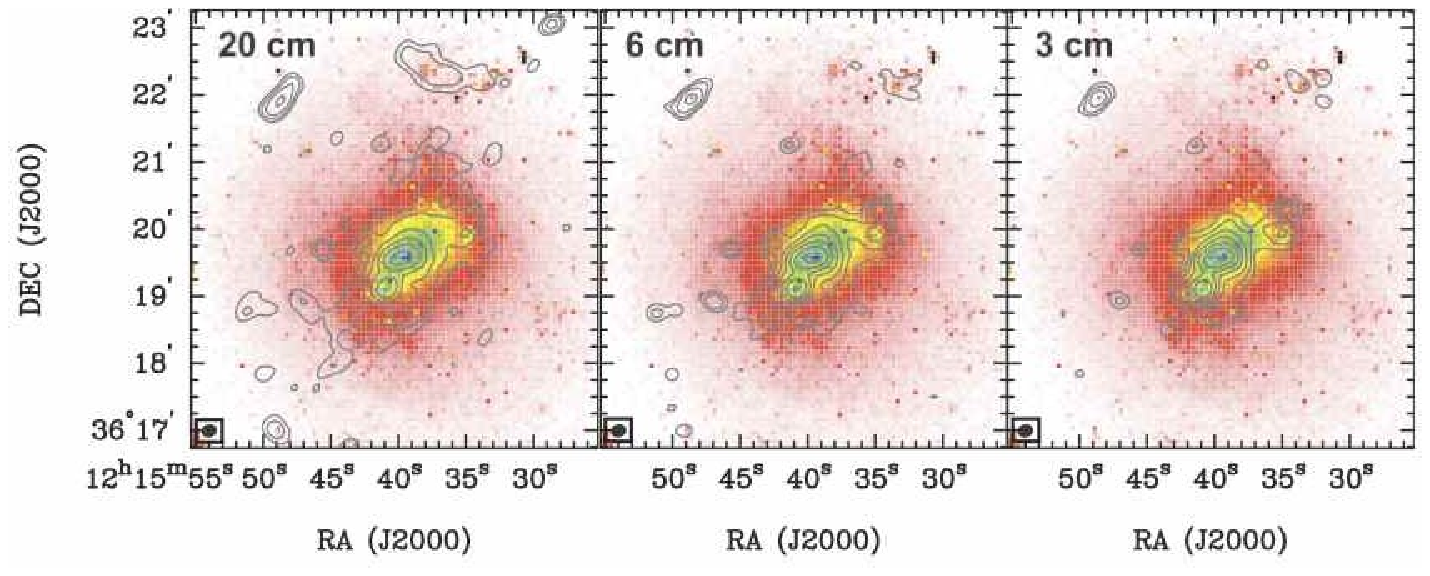}
\caption[Comparison of optical and radio continuum emission of NGC
4214]{ Contours of total intensity radio continuum emission at 20\,cm
  (left panel), 6\,cm (middle panel), and 3\,cm (right panel) overlaid
  on an optical V band image of NGC 4214 kindly provided by Deidre
  Hunter \citep{ch3:hunter2006ubv}.  The contours are 3, 6, 12, 24,
  48, 96, 192, and 384 times the 1$\sigma$ noise level given in
  Table~\ref{tab:final_image_summary}. The beam is boxed in the lower
  left corner of each panel.}
\label{ch3:fig:optv}
\end{figure}

In Figures~\ref{fig:opt_lband}, \ref{ch3:fig:halpha}, and
\ref{ch3:fig:optv}, we compare the distribution of the radio continuum
emission from NGC 4214 with the distribution of its optical and \ha\
emission. Figure~\ref{fig:opt_lband} is a three-color image relating
the \ha\ (red), optical U band (blue), and optical V band (green) to
the 20\,cm radio continuum emission. In Figures \ref{ch3:fig:halpha}
and \ref{ch3:fig:optv}, we provide more detailed comparisons of the
20\,cm, 6\,cm, and 3\,cm emission with the \ha\ and optical V band,
respectively.

From Figure~\ref{ch3:fig:halpha}, we can see that the radio continuum
at 3\,cm closely follows the distribution of the \ha\ emission. This
correspondence shows that the 3\,cm radio continuum emission has a
large contribution from thermal emission. As the wavelength of our
observations increases, we see more diffuse radio continuum
emission. In the northern half of the galaxy, there is diffuse
emission that corresponds to the faint \ha\ filaments at 12$^{\rm
h}$15$^{\rm m}$37$^{\rm s}$, 36$\arcdeg$20$\arcmin$45$\arcsec$. In the
south, there is some diffuse emission that does not closely follow the
\ha\ distribution at 12$^{\rm h}$15$^{\rm m}$45$^{\rm s}$,
36$\arcdeg$18$\arcmin$00$\arcsec$ and 12$^{\rm h}$15$^{\rm m}$47$^{\rm
s}$, 36$\arcdeg$18$\arcmin$45$\arcsec$; these regions have a large
contribution from synchrotron emission.  The western tip of the
diffuse synchrotron region ($12^{\rm h}15^{\rm m}38^{\rm s},
36^\circ22\arcmin30\arcsec$) is associated with \ha\ emission and some
faint 6 and 3\,cm emission. This part of the region is likely
dominated by thermal emission. The remainder of the diffuse
synchrotron region does not have any corresponding \ha\ emission or
any appreciable emission at 3 or 6\,cm, showing that is it dominated
by non-thermal emission.

Figure~\ref{ch3:fig:optv} shows that the radio continuum emission is
confined to the central regions of this galaxy. We see that the radio
continuum emission, the \ha\ emission, and the stellar emission are
all coincident in a bar-shaped region at the center of the galaxy.
The diffuse synchrotron region ($12^{\rm h}15^{\rm m}38^{\rm s},
36^\circ22\arcmin30\arcsec$) has stars associated with it. This
correlation leads us to believe that this feature is an older star
formation region. The noticeably resolved, optical object in the
center of the region is green in Figure \ref{fig:opt_lband}, which
implies its emission is dominated by optical V band emission and thus
belongs to an older stellar population. The expected lifetime for the
free-free emission is about 10~Myr (approximately the lifetime of the
most massive stars).  The synchrotron emission lifetime is set by the
confinement time for cosmic rays, i.e., the time it takes for cosmic
rays to diffuse out of a galaxy.\footnote{Small distortions in the
galactic magnetic field randomly scatter cosmic rays. See
\citet{2009arXiv0905.3071D} for a review of the relevant physics.}
For our Galaxy, this timescale is on the order of 10~Myr. In
Section~\ref{sec:cosic-ray-diffusion}, we estimate that this time
scale is also approximately 10~Myr for NGC 4214 (although see the
cavaets discussed in Section~\ref{sec:cosic-ray-diffusion}).  There is
a delay expected between the synchrotron emission and the free-free
emission since the first supernovae do not appear until 3.5~Myr after
the initial burst of star formation. This argument suggests that there
can be synchrotron emission in a region after the free-free emission
has dissipated. Note that in this picture, we are assuming that the
cosmic rays have not diffused very far from their sources on
timescales of 10 Myr.

The timescales for free-free and synchrotron emission allow us to
estimate the age of the star formation episode that formed the diffuse
synchrotron region. The eastern half of the region has formed stars
greater than 10~Myr ago so the free-free emission has dissipated, but
less than 13.5~Myr ago, so the cosmic rays are still confined to the galaxy
and producing synchrotron emission. The western half has formed stars
within 10~Myr and thus still has both synchrotron and free-free
emission.  The lack of emission in diffuse synchrotron region at 6 and
3\,cm is due to the decrease in synchrotron intensity as the
wavelength decreases (i.e., as the frequency increases). Assuming a
spectral index of \mbox{--0.7} and using the intensity of the diffuse
synchrotron emission at 20\,cm, the synchrotron emission from this
patch would be below the 3$\sigma$ detection limit at 6\,cm.

\subsection{Fraction of Thermal Emission} \label{sec:fract-therm-emiss}

To determine the magnetic field strength, we need to determine how
much of the radio continuum emission is due to non-thermal
(synchrotron) emission and how much is due to thermal (bremsstrahlung)
emission. Since the electrons that produce the thermal emission in the
radio are the same electrons that produce \ha\ in the optical, we use
a flux calibrated \ha\ image kindly provided by D.\ Hunter
\citep{ch3:hunter2004} to estimate the thermal contribution to the
radio emission. We use the approximate expression derived by
\citet{ch3:2004ApJ...606..853H} to convert the \ha\ fluxes to radio
continuum fluxes
\begin{equation} \label{ch3:eq:fth}
\left( \frac{f_\nu} {\rm{mJy}} \right) = 1.16 \left( 1 +
\frac{n(He^+)}{n(H^+)} \right) \left( \frac{T}{10^4
  \ \rm{K}}\right)^{0.617} \left( \frac{\nu} {\rm{GHz}} \right)^{-0.1}
\left( \frac{f_{H\alpha}} {10^{-12} \ \rm{erg \ cm^{-2} \ s^{-1}} }
\right)
\end{equation}
where $f_\nu$ is the radio continuum flux, $n(He^+)/n(H^+)$ is the
ratio of ionized helium to ionized hydrogen, $T$ is the temperature of
the region, $\nu$ is the frequency of the radio continuum
observations, and $f_{H\alpha}$ is the \ha\ flux. These expressions are
valid for temperatures between 10,000 K and 20,000~K and densities
between 100~cm$^{-3}$ and 1000~cm$^{-3}$.   We use a value of 0.086 for $n(He^+)/n(H^+)$
and a temperature of 10,000~K \citep{ch3:1996ApJ...471..211K}. The
synchrotron flux is then
\begin{equation}
F_{nth} = F_{tot}  \left( 1 - \frac{F_{th}}{F_{tot}}\right)
\end{equation}
where $F_{nth}$ is the flux of the synchrotron (non-thermal) emission,
$F_{tot}$ is the total flux observed, and $F_{th}$ is the thermal flux
calculated using Equation~(\ref{ch3:eq:fth}).  Any values of
$F_{th}/F_{tot}$ greater than one or below zero are set to one and
zero, respectively. In practice, this affects only a few edge pixels.

We correct the \ha\ flux for foreground extinction using the value
derived from the maps of \citet{ch3:1998ApJ...500..525S}: $A_R \sim
A_{H\alpha} \sim 0.058$. We do not correct for the internal extinction
of NGC 4214, which varies as a function of position in the galaxy
\citep{ch3:1998A&A...329..409M}. \citet{2008arXiv0808.1953L} find that
the internal extinction of NGC 4214 for \ha\ is between 0.3 and 0.7
magnitudes. The values of high internal extinction in this galaxy are
slightly offset from the peaks in the \ha\ emission. Due to our low
resolution ($\sim 14\arcsec$), both the peaks of thermal radio/\ha\
emission and the regions of high extinction fall within the same
beam. Not correcting for the effect of internal extinction could us to
underestimate the flux of the free-free emission by a factor of 1.3 to
1.9 and thus to overestimate the synchrotron flux by factors of 1.25
to 2 for regions that have 0.3 magnitudes of extinction and by factors
2 to 3 for regions that have 0.7 magnitudes of extinction. The lower
limits in both cases refer to the flux peaks. Given the value of
$\alpha$ used in our calculation of the magnetic field (see
Section~\ref{ch3:sec:magn-field-strength} for details), this will
change our magnetic field estimate by 9 to 40\%, with the mean change
closer to 9\%. Correcting the free-free emission also has an effect on
the measured polarization of the emission. If the \ha\ data is not
corrected for extinction, the flux of the free-free emission will be
an underestimate, so the intensity of the synchrotron will be an
overestimate. Then the percent polarization of the synchrotron
emission, which is related to the ratio of the uniform field to the
random field, will be an underestimate.

\begin{figure}
\centering
\includegraphics[scale=1.0]{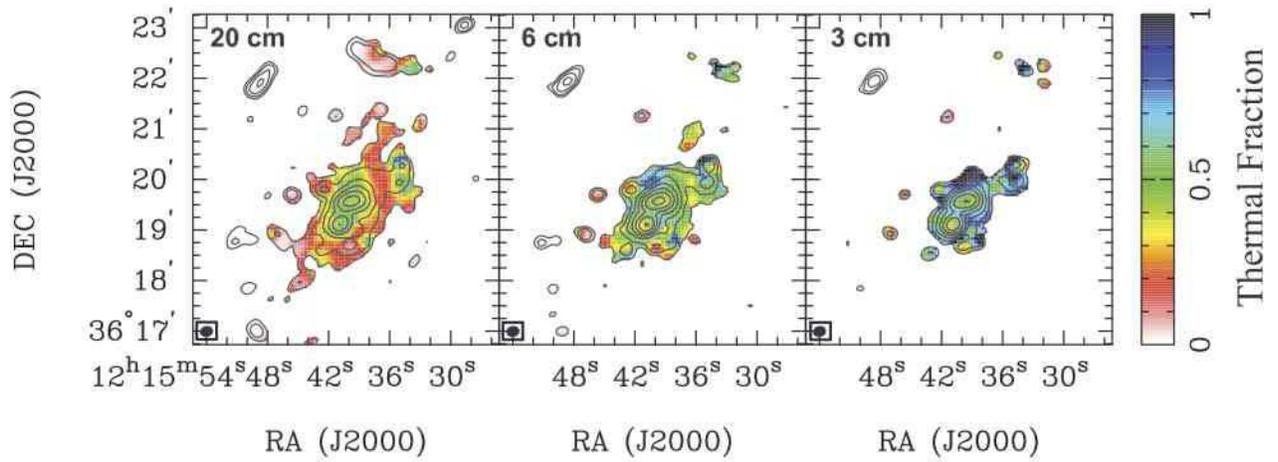}
\caption[Estimated thermal fraction for NGC 4214]{Estimated thermal
  fraction of NGC 4214 with contours of total intensity radio
  continuum emission at 20\,cm (left panel), 6\,cm (middle panel), and
  3\,cm (right panel). The contours are 3, 6, 12, 24, 48, 96, 192, and
  384 times the 1$\sigma$ noise level given in
  Table~\ref{tab:final_image_summary}. The beam is boxed in the lower
  left corner of each panel. }
\label{ch3:fig:thermal_fraction}
\end{figure}

Our thermal flux estimates as a function of position in NGC 4214 are
shown in Figure~\ref{ch3:fig:thermal_fraction}.

\subsection{Magnetic Field Strength} \label{ch3:sec:magn-field-strength}

We can calculate the magnetic field strength using the estimated
synchrotron flux from Section~\ref{sec:fract-therm-emiss}. We use the
revised equipartition estimate of the magnetic field strength given in
\citet{ch3:2005AN....326..414B}
\begin{equation}
  \label{ch3:eq:beq}
  B_{t} = \left[ \frac{ 4 \pi (1 - 2 \alpha) ( K_0 + 1) I_\nu
      E_p^{1 + 2\alpha} (\nu/2c_1)^{-\alpha} }
{ (-2\alpha-1)c_2(\alpha) l c_4} \right]^{1/(3 - \alpha)}
\end{equation}
where $B_{t}$ is the total equipartition magnetic field in G, $\alpha$
is the spectral index of the emission\footnote{ We use the opposite
sign convention for $\alpha$ as \citet{ch3:2005AN....326..414B}.},
$K_0$ is the number density ratio of the protons to the electrons,
$E_p$ is the rest energy of the proton, $c_1$ is a constant equal to
$6.26428\times 10^{18} \ \rm{erg}^{-2} \ \rm{s}^{-1} \ \rm{G}^{-1}$,
$c_2$ is a constant tabulated on page 232 of
\citet{ch3:1970ranp.book.....P} as $c_5$, $l$ is the line of sight
through the galaxy, and $c_4$ is a function correcting for the
inclination of the region with respect to the sky.  Since our
polarization observations shown in
Section~\ref{sec:polarized-emission} do not reveal a large-scale
uniform field, we use the version of $c_4$ for the case of a
completely random, isotropic field
\begin{equation}
c_4 = \left( \frac{2}{3} \right)^{(1-\alpha)/2}.
\end{equation}
In this case, $c_4$ does not explicitly depend on the inclination, but
instead represents an average over all possible orientations for the
random field.

We need to carefully consider which frequency has the best estimate of
the synchrotron flux. In Section~\ref{sec:total-intensity}, we see
hints that the radio continuum spectrum of NGC 4214 may be affected by
either free-free absorption of synchrotron emission and/or synchrotron
losses, although we do not have adequate data at higher and lower
frequencies to constrain either model. Free-free absorption of
synchrotron emission would cause the 20\,cm synchrotron flux to be
underestimated. Cosmic rays also lose energy via synchrotron emission
if they are not reaccelerated.  This effects the higher energy cosmic
rays more than the lower energy cosmic rays and would thus cause the
3\,cm synchrotron flux to be underestimated. Therefore, the 6\,cm
image represents the best wavelength for us to determine the magnetic
field strength of NGC 4214. We use a typical value for $\alpha$ of
\mbox{--0.7}. Given the uncertainties in the mechanisms responsible
for shaping the radio continuum spectrum of NGC 4214, it is better to
use a typical value than a value derived from the data. We note that
the spectral index between the 20\,cm and 3\,cm synchrotron fluxes for
our data is --0.6, which is very close to the typical value. 

We can use the following expression from
\citet{ch3:2001AJ....122.3070O} to estimate the scale height of NGC
4214's disk from neutral hydrogen observations and obtain the line of
sight distance through NGC 4214's disk
\begin{equation}
h \left[ {\rm pc} \right] = 
5.79 \times 10^{21} 
\left( \frac{\sigma_{gas}}{\kms} \right)^2 
\left( \frac{N_{HI}}{{\rm cm}^{-2}}\right)^{-1} 
\left( \frac{\rho_{HI}}{\rho_{tot}} \right)    
\end{equation}
where $h$ is the scale height, $\sigma_{gas}$ is the velocity
dispersion of the neutral gas, $N_{HI}$ is the column density of the
neutral gas, and $\rho_{HI}/\rho_{tot}$ is the ratio of the density of
neutral gas to the total density. From
\citet{ch3:1979MNRAS.188..765A}, we know that $M_{HI}/M_{tot}$ is 0.3,
so since $ M_{HI}/M_{tot} \sim \rho_{HI}/\rho_{tot} =
0.3$. \citet{ch3:1979MNRAS.188..765A} also gives us a column density
for the neutral gas of $8 \times 10^{20} \ {\rm cm^{-2}}$ and a
velocity dispersion of 10~\kms. These values give a scale height for
this disk of 200~pc, which agrees with the estimate of the disk scale
height by \citet{ch3:1999A&A...343...64M}. Given the uncertain but
close to face-on inclination of NGC 4214 we take $i=30^\circ$.

The scale height of the magnetic field may be larger than the scale
height of the HI. In the Milky Way, the scale height of the HI is
1.6~kpc \citep{2011A&A...525A.134M}, while the scale height of the
synchrotron emission is 4.6~kpc at the solar circle
\citep{2001RvMP...73.1031F}. Unfortunately, the face-on orientation of
NGC 4214 and the lack of information about its overall field topology
prevent us from determining the scale height of the synchrotron
emission from first principles. We can estimate the effect of a much
larger magnetic field scale height on the magnetic field strength
estimates for NGC 4214 using the difference between the Milky Way HI
and synchrotron scale height and Equation~(\ref{ch3:eq:beq}). If the
synchrotron scale height of NGC 4214 is three times larger than the HI
scale height, then the magnetic field strength estimate above will be
1.3 times bigger than the actual magnetic field strength.

\begin{figure}
\centering
\includegraphics[scale=0.8]{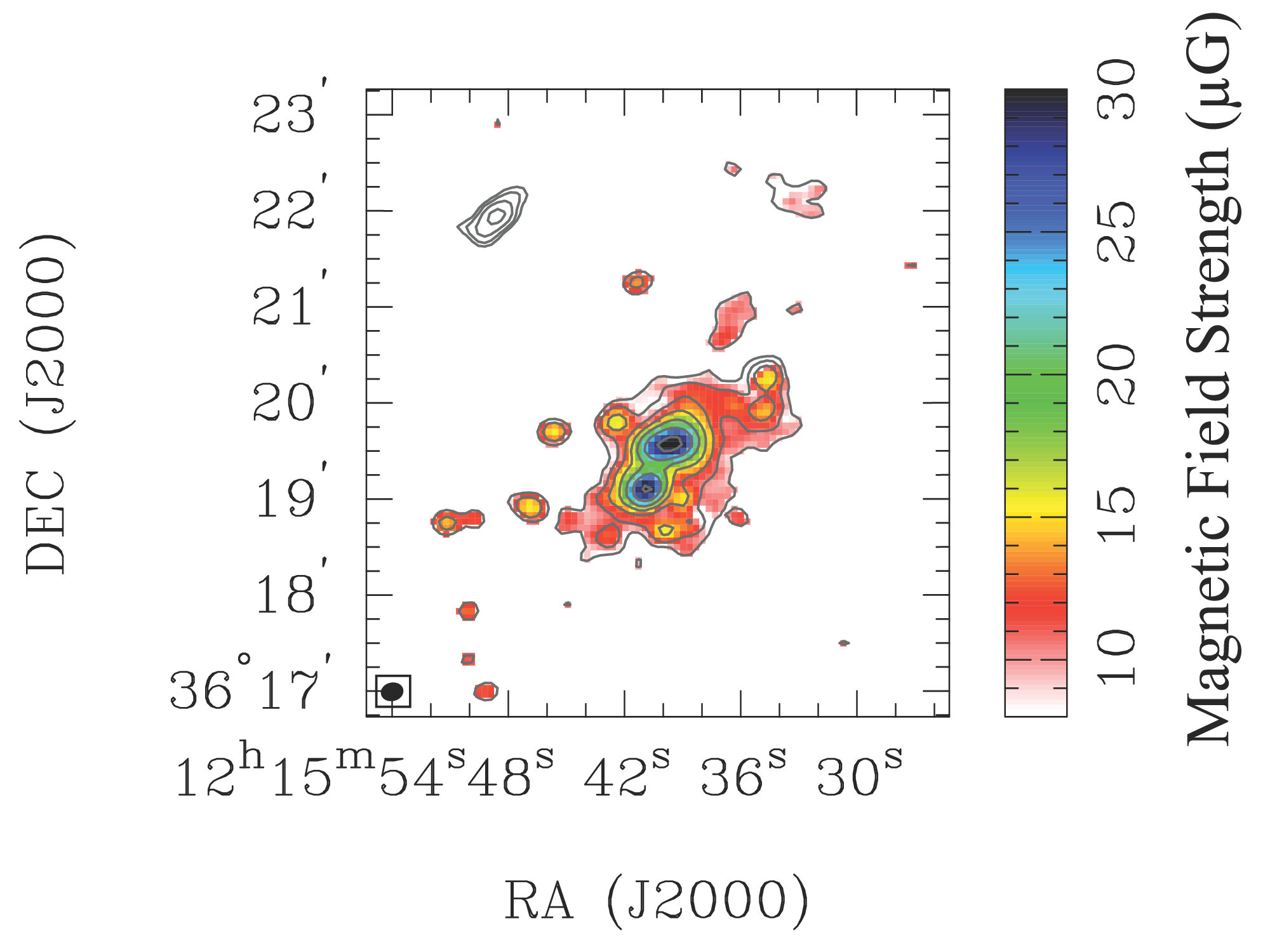}
\caption[Estimated magnetic field strength of NGC 4214]{ Magnetic
field strength (in $\mu\rm{G}$) calculated using the estimated
synchrotron emission at 6\,cm. The magnetic field has not been
calculated for the background AGN at ($12^{\rm h}15^{\rm m}48.9^{\rm
s}, 36^\circ21\arcmin54.10\arcsec$) since it does not belong to NGC
4214. The beam is boxed in the lower left.}
\label{ch3:fig:bfield_tot_lband}
\end{figure}

Our estimate of the magnetic field strength in NGC 4214 is given in
Figure~\ref{ch3:fig:bfield_tot_lband}. The magnetic field strength
peaks near the center of the galaxy with a value of 30~\uG\ and
decreases to a value of 10~\uG\ at the edge of the galaxy. To get an
idea of the uncertainties in these estimates, we have varied $K_0$ to
40, $\alpha$ by 0.1, changed the estimated synchrotron emission by a
factor of 2, changed the line of sight distance through the galaxy by
20\%, and changed the distance by 20\%. With these uncertainities in
the initial parameters, we get an uncertainty for the central magnetic
field strength of 9.5~\uG\ and an uncertainty for the magnetic field
strength on the edges of 3~\uG.

\subsection{Importance of the Magnetic Field in the ISM of NGC 4214} \label{sec:import-magn-field}

To determine the relative importance of the magnetic field in the
larger context of the interstellar medium of NGC 4214, we calculate
total magnetic pressure as well as the magnetic pressure in the inner
and outer regions of NGC 4214 and compared this pressure to the
thermal pressure from the hot ionized gas, the thermal and turbulent
pressure in HII regions, and the gravitational ``pressure''. These
values, along with the values and equations used to calculate them,
are given in Table~\ref{tab:ism_pressures}.

From these estimates, we can see that the total hot gas pressure, the
total magnetic pressure, and the gravitational ``pressure'' are the
same order of magnitude, with the hot gas having the largest pressure
and the magnetic field the lowest.\footnote{Since the gravitational
pressure was estimated using the Virial Theorem, it should only be
compared to the pressures for the entire system, not the pressures for
individual regions.} The thermal and turbulent pressures
of the HII regions, however, dominate the pressure balance in the
central star forming regions (although not the galaxy as a
whole). \citet{ch3:2001ApJ...555..758W} found that the energy in the
ionized gas was comparable to that injected via stellar winds. The ISM
pressure balance of NGC 4214 is similar to that of NGC 1569
\citep{2010ApJ...712..536K}; the turbulent and thermal pressures of
the HII regions dominate the pressure balance of NGC 1569 in the
central star forming regions, while the magnetic field, hot gas, and
gravity are approximately equal. Although the magnetic field strength
in both galaxies is comparable to that of larger spiral galaxies,
neither NGC 4214 nor NGC 1569 appear to be magnetically overpressured
compared to their gravitational potential. The magnetic field of NGC
4214 may play a role in shaping any outflow of gas from this galaxy as
in NGC 1569 \citep{2010ApJ...712..536K}, but given the orientation of
NGC 4214, we are not able to determine this with our observations (see
Section~\ref{sec:structure-halo-ngc}).

\subsection{The Structure of the Magnetic Field of NGC 4214} \label{sec:struct-magn-field}

\subsubsection{Uniform Field Using Polarization Information} \label{sec:uniform-field-using}

Using our data, we can calculate an upper limit on the uniform
magnetic field strength in NGC 4214. One can approximate the percent
polarization as
\begin{equation} \label{eq:frac_pol}
p = \frac{P}{I} = p_0 \frac{ B_{u,\perp}^2 } {B_{t,\perp}^2} 
\end{equation} 
where $p$ is the percent polarization, $P$ is the intensity of the
polarized emission, $I$ is total intensity of the synchrotron
emission, $p_0$ is the intrinsic polarization of synchrotron emission
($\sim 0.75$), $B_{u,\perp}$ is the uniform magnetic field in the
plane of the sky, and $B_{t,\perp}$ is the total magnetic field in the
plane of the sky
\citep{1966MNRAS.133...67B,ch3:2003A&A...411...99B}. For NGC 4214, we
calculated the total field in Section
\ref{ch3:sec:magn-field-strength} assuming that the field was entirely
random, so Equation~(\ref{eq:frac_pol}) becomes
\begin{equation} \label{eq:frac_pol_ran}
  p = p_0 \frac{ B_{u,\perp}^2 } { \frac{2}{3} B_t^2 }.
\end{equation}

Assuming a value of 40\% thermal emission at 6\,cm (see
Section~\ref{sec:fract-therm-emiss}), we would detect at the 3$\sigma$
level 30\% polarization at 12 times the noise level (third contour in
Figure~\ref{ch3:fig:polarized_intensity}). In
Section~\ref{ch3:sec:magn-field-strength}, we estimate the total
magnetic field strength at the 3rd contour to be 15~\uG.\footnote{
  Recall that this field strength was derived in
  Section~\ref{ch3:sec:magn-field-strength} assuming that the field
  was entirely random.} Using Equation~(\ref{eq:frac_pol_ran}), we
then get an upper limit on the strength of the uniform component of
the magnetic field in the plane of the sky of 8~\uG. The large thermal
component of the radio continuum emission of NGC 4214 reduces our
sensitivity to the uniform field. If the radio continuum emission in
NGC 4214 was all synchrotron, our upper limit on the uniform magnetic
field strength would be 6~\uG.

From this analysis, we conclude that we do not detect a uniform
component of the disk magnetic field of NGC 4214 greater than 8~\uG\
on size scales larger than our resolution (200~pc). The disk of NGC
4214 may have a uniform field, but it would have to be weaker than
8~\uG. There could be structure in the magnetic field on size scales
smaller than 200~pc, but our observations cannot resolve these
structures.

\subsubsection{Could the Shape of the Diffuse Synchrotron Region Be
  Due to a Uniform Field?} \label{sec:unif-field-estim}

As a star-forming region ages, the relativistic electrons diffuse away
from their birthplaces along magnetic field lines. In a completely
random field, one would expect that the electrons would diffuse
outward isotropically. If there is a uniform component to the field,
however, the relativistic electrons will preferentially diffuse along
the field lines rather than across them. The diffuse synchrotron
region in NGC 4214 is elongated along a position angle of about 60
degrees, which suggests that there is a uniform magnetic field
component in that direction.

We can use the distance between the stars and the end of the region
(490~pc) and the probable age of the region (10~Myr) to estimate the
speed at which the relativistic electrons diffused. This gives a lower
limit on the diffusion speed of $4.7 \times 10^6$~\cms. Assuming that
relativistic electrons are moving at the \alfven\ speed ($v_A$)
\begin{equation}
v_A^2 = \frac{B^2}{4 \pi \rho} 
\end{equation}
where $B$ is the magnetic field strength, $\rho$ is the density, and
all quantities are in cgs units, if we have an estimate of the density
in the region then we can get a lower limit for the strength of the
magnetic field responsible for diffusing the cosmic rays.  The speed
of the cosmic rays is coupled to the density of the thermal electrons,
so we can use the \ha\ image described above to get an estimate of the
density of the thermal electrons in the diffuse synchrotron region. In
the diffuse synchrotron region there is $4.11 \times 10^{-6} \ {\rm
erg \ cm^{-2} \ s^{-1} \ sterad^{-1}}$ of emission. From
\citet{1978ppim.book.....S},
\begin{equation}
{\rm EM} \left[ {\rm pc \, cm^{-6}} \right] = \frac{\int I_\nu \,
  \rm{d} \nu \lambda}{2.46 \times 10^{17} \, h \, c \, \alpha_{3,2} }
\end{equation}
where EM is the emission measure, $I_\nu$ is the specific intensity,
$\alpha_{3,2}$ is the production coefficient for the \ha\ line, and
all values (unless noted) are in cgs. We use a value of $\alpha_{3,2}$
of $11.7 \times 10^{-14}$ which is valid for 10,000~K. This gives us
an emission measure for the region of 47~pc~cm$^{-6}$. If we assume
that NGC 4214 is nearly face-on, we use the line of sight distance
through the disk derived in Section~\ref{ch3:sec:magn-field-strength}
to convert the emission measure to a number density of thermal
electrons. This calculation gives a number density for the electrons
of 0.34~cm$^{-3}$. Assuming that the ionized gas has the same number
of electrons and protons gives a density for the ionized gas of $5.7
\times 10^{-25} \ {\rm g \, cm^{-3}}$. The lower limit on the total
magnetic field strength can then be derived using
\begin{equation}
B = v_A \sqrt{4 \pi \rho}.
\end{equation}
Using our values of $v_A$ and $\rho$, the lower limit on the total
magnetic field strength in the diffuse synchrotron region is 13~\uG.
This field strength is the same as the field strength derived for the
diffuse region from the 20\,cm data using the assumptions described in
Section~\ref{ch3:sec:magn-field-strength}. At 20\,cm, we are sensitive
to polarizations greater than 26\% in the diffuse synchrotron
region. Using the magnetic field strength determined from the
synchrotron emission (13~\uG), we get an upper limit on the uniform
field strength of 7.6~\uG.

We can use the ratio of the length of the region to the width of the
region to derive the ratio of the uniform magnetic field strength
($B_u$) to the random magnetic field strength ($B_r$) and compare it
to the limits on $B_u/B_r$ from our data. Let us take $y$ to be the
direction along the diffuse synchrotron region and $x$ to be the
direction across it, the equation of the magnetic field to be ${\bf B}
= \hat{y} B_y + \hat{x} B_x$, $X$ to be half the total width of the
region, and $Y$ to be half the total length of the region. $B_y$ and
$B_x$ are both constant in a correlation length $l$, but $B_x$ can be
positive or negative. In this picture, $B_y$ corresponds to the
uniform component and $B_x$ corresponds to the random component. Then
the equation of a field line is
\begin{equation}
  \frac{\rm{d}y}{\rm{d}x} = \frac{B_y}{B_x}.
\end{equation}
The displacement in $y$ when the field line has gone a distance $l$ in
$x$ is 
\begin{equation}
  \Delta y = \frac{l B_y}{B_x}.
\end{equation}
The cosmic rays stream at the \alfven\ speed, so the time to travel
the distance $l$ is $\tau = l / v_{Ax}$, where $v_{Ax} = B_x / (4 \pi
\rho)^{1/2}$. After a time $t$, the cosmic rays have taken $N =
t/\tau$ random steps of length $l$ in the $x$ direction , so the $rms$
displacement of the cosmic rays, $X$, is $l N^{1/2} = (l v_{Ax}
t)^{1/2}$. The displacement in the $y$ direction is just $N \Delta y =
N l B_y / B_x$. Then the elongation of the cloud is
\begin{equation}
\frac{Y}{X} = \left( \frac{ v_{Ax} t }{l} \right)^{1/2}
\frac{B_y}{B_x} = N^{1/2} \frac{B_y}{B_x} .
\end{equation}
We can estimate $N$ from the strength of the random field derived
above from our data (12.9~\uG), the density of the region
(0.34~cm$^{-3}$), the age of the region (10~Myr), and the outer length
scale for turbulence \citep[90~pc in the
LMC;][]{ch3:gaensler2005}. This gives a value for $N$ of approximately
6. The measured values of X and Y are 235~pc and 491~pc, respectively,
which gives a value for $Y/X$ of 2.1. With our estimate of $N$, the
ratio $B_y/B_x = B_u/B_r$ is 0.85. From our data, the ratio of
$B_u/B_r$ allowed by our data is 0.69, which is close enough to the
ratio estimated from the size of the region to make our scenario that
the diffuse synchrotron region is elongated by a uniform field
plausible.

\subsubsection{Possible Structure in the Halo of NGC 4214} \label{sec:structure-halo-ngc}

The analysis in Sections \ref{sec:uniform-field-using} and
\ref{sec:unif-field-estim} only applies to the structure of the
magnetic field in the disk of NGC 4214. The nearly face-on orientation
of NGC 4214 does not allow us to determine whether NGC 4214 has a
uniform magnetic field oriented along a galactic outflow like that
seen in NGC 1569 \citep{2010ApJ...712..536K}. From estimates of the
mechanical energy in the wind ($L_{mech}\sim 10^{40} - 10^{41} \
\rm{erg \ s}^{-1}$), it appears that it is possible for NGC 4214 to
experience a blow-out \citep{ch3:ott05II}. Therefore, NGC 4214 might
have magnetic field structure perpendicular to its disk in its
halo. Detection of this halo field may be possible using the dense
grid of background rotation measures produced by the Square Kilometer
Array \citep{2004NewAR..48.1289B}.

We would like to note that while in principle the rotation measure of
the background source to the northeast of the main body of emission
could be used to estimate the magnetic field direction and strength
along the line of sight, i.e., through the disk, unfortunately there
is no \ha\ emission in the region of the galaxy associated with the
background source, so we would be unable to determine a field
strength. In addition, there are no other nearby background sources to
allow us to estimate an average rotation measure for background
sources in that region of the sky.

\subsection{Cosmic Ray Lifetimes} \label{sec:cosic-ray-diffusion}

Using the arguments from Section~\ref{sec:unif-field-estim}, we can
also estimate a timescale for cosmic ray confinement in NGC
4214. Assuming that a supernova goes off at the midplane of the disk,
the cosmic rays would have to travel 200~pc to to escape the
disk. Since cosmic rays diffuse at the \alfven\ velocity, the total
displacement of the cosmic rays, $L$, in a time $t$ is
\begin{equation}
L = (l v_A t)^{1/2} =\frac{ ( l B t)^{1/2} } { (4 \pi \rho )^{1/4}}.
\end{equation}
The timescale for the cosmic rays to escape the disk is then
\begin{equation}
t = \frac{X^2 (4 \pi n_e m_H)^{1/2}}{l B}, 
\end{equation}
where $n_e$ is the number density of electrons and $m_H$ is the mass
of a hydrogen atom. We estimate $n_e$ to be 1~$\rm{cm^{-3}}$, $B$ to
be 20~\uG, and take $l$ to be the outer length scale of the
turbulence, which \citet{ch3:gaensler2005} found to be 90~pc in the
LMC. Using these parameters, the timescale for the diffusion of cosmic
rays out of the disk of NGC 4214 is approximately 10~Myr -- comparable
to the diffusion time in the Milky Way. If the scale height of the
synchrotron emission is much larger, as discussed at the end of
Section~\ref{ch3:sec:magn-field-strength}, this timescale could be as
long as 67~Myr. NGC 4214 may have a wind in its central portion
\citet{ch3:ott05II} which would shorten the cosmic ray confinement
time \citep{2008ApJ...674..258E,2010ApJ...711...13E}. Detailed
modeling of the effect of a central wind on the cosmic ray confinement
time in NGC 4214 is beyond the scope of the paper.

Synchrotron losses will further limit the lifetime of synchrotron
emission. To see if the synchrotron loss timescale will dominate over
the cosmic ray confinement time, we calculate the lifetime of cosmic
ray electrons due to synchrotron losses. The power $P$ in erg~s$^{-1}$
radiated by synchrotron emission is
\begin{equation} \label{eq:synchtron_losses}
P = \frac{4}{3} \sigma_T c \beta^2 \gamma^2 \frac{B^2}{8\pi}
\end{equation}
where $\sigma_T$ is the Thomson cross-section, $c$ is the speed of
light, $\beta$ is $v/c$, and $\gamma$ is $(1-\beta^2)^{-1/2}$. Since
cosmic rays are traveling very close to the speed of light, we can
assume $\beta \approx 1$.
Assuming that all the emission is produced at the critical frequency
\begin{equation} \label{eq:critical_freq}
\nu_c = \frac{3}{4 \pi} \gamma^2 \frac{q B}{mc} \sin{\alpha}
\end{equation}
and using the relation $E=\gamma m c^2$, we find that the energy of
the electron at the critical frequency is
\begin{equation} \label{eq:electron_energy}
E = \left( 
\frac{4 \pi m_e^3 c^5 \nu_c}{3 q B}
\right)^{1/2}
\end{equation}
where the factor of $\sin{\alpha}$ has been ignored since it is a
factor on the order of one. Dividing the result of Equation
(\ref{eq:electron_energy}) by Equation (\ref{eq:synchtron_losses})
gives an estimate for the synchrotron loss timescale.  For emission at
20\,cm (1.4~GHz) and a magnetic field strength of 10~\uG\ (the field
strength in the diffuse synchrotron region), the synchrotron loss
timescale for a cosmic ray electron is 44~Myr. At 6\,cm (8.5~GHz), the
lifetime of the cosmic rays due to synchrotron losses is shorter -- 18
Myr. Therefore, at 20\,cm in the diffuse synchrotron region, the
confinement time is shorter than the synchrotron loss timescale and
thus the confinement time dominates. (Although if the synchrotron
scale height is 600~pc rather than 200~pc, the synchrotron loss time
will dominate.) At lower frequencies and larger magnetic fields, the
synchrotron loss time scale decreases so that in the center of the
galaxy with a magnetic field of 30~\uG\ and at 3\,cm, the synchrotron
loss timescale may be as short at 6~Myr.

\subsection{Generating the Magnetic Field} \label{sec:gener-large-scale}

 To evaluate the possibility of an $\alpha-\omega$ dynamo operating in
NGC 4214, we compare the shear and turbulence in NGC 4214 to that of a
larger galaxy, where an $\alpha-\omega$ dynamo may be operating: the
Milky Way.

\begin{figure}
\centering
\includegraphics[scale=0.8]{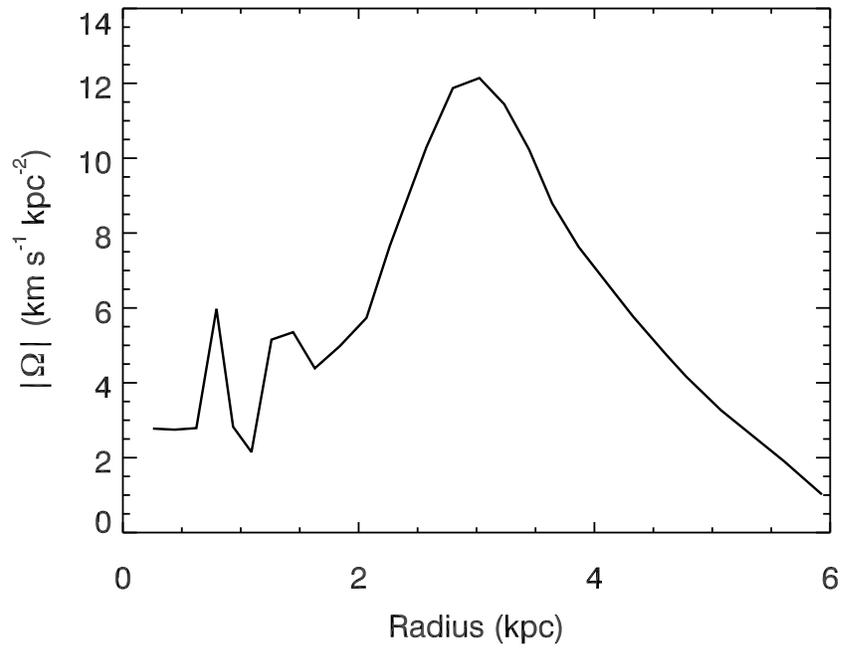}
\caption{Absolute value of the shear calculated using the neutral
  hydrogen rotation curve from \citet{ch3:1979MNRAS.188..765A},
  corrected for the distance adopted in this paper and for the
  inclination of NGC 4214, as a function of radius.}
\label{fig:n4214_shear_vs_radius}
\end{figure}

First, we evaluate the effect of differential rotation by calculating
the shear in both systems. Shear is defined as ${\rm d} \Omega /{\rm
d} r$, which for a flat rotation curve is just $v/r^2$. At the solar
radius (8~kpc), the rotation speed is 220~\kms, so the shear in the
Milky Way is about 3.4~${\rm km \, s^{-1} \, kpc^{-2}}$. In
Figure~\ref{fig:n4214_shear_vs_radius}, we show the absolute value of
the shear in NGC 4214 as a function of radius.  To calculate the
shear, we used the neutral hydrogen rotation curve for NGC 4214 given
in \citet{ch3:1979MNRAS.188..765A}, corrected for the distance and
inclination used in this paper. The shear in NGC 4214 varies
significantly below a radius of 2~kpc, but the average value at these
radii is close to the Milky Way value. At a radius of 2~kpc, the shear
begins to increase, reaching a maximum of 12~${\rm km \, s^{-1} \,
kpc^{-2}}$ at a radius of about 3~kpc. The shear then begins to
decrease and by 5~kpc is at or below the shear estimated at the solar
radius in the Milky Way. We would note that the diffuse synchrotron
region discussed in Section~\ref{sec:unif-field-estim} is located in
the region of maximum shear about 2.5~kpc from the center of NGC
4214. According to Rayleigh's Criterion, the quantity $\Omega r^2$
must increase as a function of $r$ for a rotation curve to be stable.
This is satisfied for the $\Omega$ shown in
Figure~\ref{fig:n4214_shear_vs_radius} everywhere except in $3 \, {\rm
kpc} < r < 5.5 \, {\rm kpc}$, but the observational uncertainties are
large enough that the departure from stability, which is slight, is
most likely due to error.

To evaluate the turbulent component of the dynamo process, we assume
that the turbulence in these galaxies is generated by supernovae and
use the supernova rate as a proxy for turbulence. In the Milky Way,
the supernova rate from type II supernovae is 27~${\rm Myr^{-1} \,
kpc^{-2}}$ \citep{2001RvMP...73.1031F}. For NGC 4214, we can determine
the supernova rate from the star formation
rate. \citet{ch3:hunter2004} determine a star formation rate per unit
area for NGC 4214 of $5.0 \times 10^3 \ {\rm M_\sun \, yr^{-1} \,
kpc^{-2}}$. Assuming a Salpeter IMF, that the range of masses for
stars is 0.8~M$_\odot$ to 150~M$_\odot$, and that only stars above
8~M$_\odot$ become Type II supernovae, we obtain a rate for the
formation of stars greater than 8~M$_\odot$ of 85~${\rm Myr^{-1} \,
kpc^{-2}}$. Assuming a massive star lifetime of 30~Myr and that the
burst of star formation in NGC 4214 began 4~Myr ago
\citep{ch3:2000AJ....120.3007M}, that gives a supernova rate of
11.3~${\rm Myr^{-1} \, kpc^{-2}}$.

Comparing the shear and the supernova rates between the Milky Way and
NGC 4214, we see that NGC 4214 has a similar level of shear, but a
lower supernova rate by a factor of 2. This rough calculation may
indicate that additional turbulence needs to be injected by supernovae
in order to provide the turbulence necessary for the shear to generate
a large-scale field. Given that the episode of star formation in NGC
4214 started relatively recently and that it seems to be continuously
producing stars, we anticipate that supernovae will be able to
generate additional turbulence as the galaxy ages.  In shear-dominated
turbulence, one expects that the field would be oriented (although not
necessarily {\em directed}) azimuthally. The region of diffuse
synchrotron emission is located near the shear maximum in NGC 4214,
which suggests that the magnetic field in this region is oriented by
the shear, but not necessarily directed as it would be in a large
scale dynamo.

Amplification by an $\alpha-\omega$ mechanism is rather slow, so
various authors \citep[e.g.][]{ch3:1992ApJ...401..137P} have proposed
mechanisms to decrease the time needed to amplify the field. If a
large-scale field could be generated for NGC 4214 using a faster
dynamo like the one proposed for NGC 4449 \citep{ch3:om2000}, it may
have not had time to develop yet or it may have been disrupted by
recent star formation. Note that we cannot probe magnetic field
structure on size scales less than 200~pc due to the resolution of our
observations.

\section{Conclusions}\label{ch3:sec:conclusions}

In this paper, we have presented the deepest radio continuum
polarization images of the irregular galaxy NGC 4214 to date. Below we
summarize the overall conclusions of our work.

The global radio continuum spectrum for NGC 4214 is convex, which
suggests that either free-free absorption of synchrotron emission
and/or free-free emission or synchrotron losses are occurring.  We fit
the global radio continuum spectrum of NGC 4214 with a single power
law with a spectral index of \mbox{--0.43} because of the offsets
between our data and the archival data, the limited wavelength
coverage of our data, and the heterogenity of the archival data. This
spectral index indicates that the radio continuum emission in NGC 4214
is a mix of thermal and non-thermal emission. The spectral indices of
the radio continuum peaks are flat, with the younger region (NGC
4214-II) having the flatter spectral index. Between the radio
continuum peaks, the spectral indices are steeper, showing that
synchrotron emission plays a more important role there. We do not
detect any significant polarized emission in NGC 4214 at size scales
greater than 200~pc and conclude that this is due to the intrinsic
nature of the galaxy rather than instrumental/imaging effects or
internal depolarization.

We find that the radio continuum emission at 3\,cm is correlated with
the \ha\ emission from this galaxy, while the radio continuum emission
at 20\,cm is more extended. The central regions of NGC 4214 are
approximately 50\% thermal at 3\,cm. North of the main body of radio
continuum emission, we detect a region of diffuse synchrotron
radiation associated with a group of stars in NGC 4214 and speculate
that this region is a relic of previous star formation. Based the
relative timescales of free-free and synchrotron emission, we estimate
an age for this region of between 10 and 13.5~Myr.

We estimate the fraction of thermal radio continuum emission in NGC
4214 using \ha\ images. Using this estimate, we calculate the
estimated intensity of the synchrotron radiation to estimate the
magnetic field strength of NGC 4214. The field strength reaches a
maximum of $30\pm 9.5$~\uG\ in the center of the galaxy and tapers off
to $10\pm3$~\uG\ at the edges. Comparing the pressures of the various
components of the ISM of NGC 4214, we find that the hot gas pressure,
the magnetic pressure, and the gravitational ``pressure'' all the same
order of magnitude. In the central star forming regions of NGC 4214,
the thermal and turbulent pressure of the HII regions dominate the
pressure balance of this galaxy. The pressure balance in NGC 4214 is
similar to that of another irregular galaxy: NGC 1569. In both
galaxies, star formation plays the dominant role in shaping the
ISM. Although the magnetic field strength in both galaxies is
comparable to that of larger spiral galaxies, neither NGC 4214 nor NGC
1569 appear to be magnetically overpressured compared to their
gravitational potential.

Using our polarization data, we place an upper limit on the uniform
magnetic field strength of 8~\uG\ on size scales greater than
200~pc. We speculate that the elongation of the diffuse synchrotron
emission is due to a uniform field component. Using simple estimates
of the \alfven\ speed and the density of the region along with upper
limits on the polarization from our data, we place an upper limit of
7.6\uG\ on the uniform field strength in this region and calculate a
total field strength of 13~\uG\ and a random field strength of
11~\uG. These limits are consistent with the observed elongation of
the diffuse synchrotron region. 

We use the simple model developed in
Section~\ref{sec:unif-field-estim} to estimate the cosmic ray
confinement time in NGC 4214 and find it to be appropriately 10~Myr,
comparable to the diffusion time in the Milky Way. Synchrotron losses
and a galactic wind may further limit the lifetimes of cosmic rays in
the galaxy.

Finally, we investigate whether an $\alpha-\omega$ dynamo could
operate in NGC 4214 by comparing its shear and supernova rate to that
of the Milky Way. We find that the shear in NGC 4214 is comparable to
the Milky Way, but that the supernova rate is half that of the Milky
Way. We suggest that there is not enough turbulence yet in the NGC
4214 to drive an $\alpha-\omega$ dynamo, but that as the star
formation in NGC 4214 progresses the additional turbulence will allow
an $\alpha-\omega$ dynamo to operate.

Of the low mass irregular galaxies with measured magnetic fields, we
find that the magnetic field of NGC 4214 most resembles that of IC 10,
i.e., strong with some evidence of uniform fields. In a future paper,
we will investigate the properties of all observed magnetic fields in
irregular galaxies and attempt to create an integrated picture of how
they are generated and sustained in these galaxies.

\acknowledgments

The authors thank the referee for their careful report. They also
gratefully acknowledge Deidre Hunter for providing the optical and
\ha\ images used in this paper. AAK would like to thank Jay Gallagher,
Sne\v{z}ana Stanimirovi\'{c}, Dan McCammon, Bryan Gaensler, Sui Ann
Mao, Jake Simon, and Phil Arras for helpful discussions. AAK
acknowledges support from a National Science Foundation Graduate
Fellowship, GBT student support awards
(GSSP07-0001,-0002,-0003,-0019), and a Wisconsin Space Grant
Consortium Graduate Fellowship. EMW acknowledges support from a grant
from the National Science Foundation (AST-0708002).

This research has made use of the NASA/IPAC Extragalactic Database
(NED) which is operated by the Jet Propulsion Laboratory, California
Institute of Technology, under contract with the National Aeronautics
and Space Administration, and of NASA's Astrophysics Data System
Bibliographic Services.

Facilities: \facility{VLA}

%----------------------------------------------------------------------
%			Appendices
%----------------------------------------------------------------------

\appendix

\section{Deriving the Gravitational
  Pressure} \label{sec:deriv-grav-press}

To derive the gravitational pressure, we start with the Virial Theorem
\begin{equation} \label{eq:virial_thm}
2 K + U = 0
\end{equation}
where $K$ is the kinetic energy and $U$ is the potential energy. Here
we assume that the density distribution of NGC 4214 can be
approximated by an isothermal sphere with density distribution
\begin{equation} \label{eq:isothermal_density}
\rho(r) = \rho_0 \left( \frac{r_0}{r} \right)^2
\end{equation}
\citep{2007AJ....133.2242K,2009A&A...505....1V}, where $\rho_0$ and
$r_0$ are the central density and core radius, $\rho$ is density, and
$r$ is radius. Note that the rotation curves of galaxies are more
typically fit with a pseudo-isothermal sphere
\begin{equation}\label{eq:pseudo_isothermal_density}
  \rho(r) = \rho_0 \left[ 1 + \left(\frac{r_0}{r}\right)^2\right]^{-1}
\end{equation}
which does not have infinite density at zero radius like
Equation~(\ref{eq:isothermal_density}) does. As $r$ becomes large, the
density from Equation~(\ref{eq:isothermal_density}) tends to the
density from Equation~(\ref{eq:pseudo_isothermal_density}). We are
measuring the potential far from the center of the galaxy, so we
assume that Equation~(\ref{eq:isothermal_density}) applies, which
yields a much more analytically tractable result. The mass interior to
radius $r$, $M(r)$, is then just
\begin{eqnarray}
M(r) & = & \int_0^r 4 \pi \rho(r^\prime) (r^\prime)^2 \, {\rm d} r^\prime \\
     & = & 4 \pi \rho_0 r_0^2 r
\end{eqnarray}
The potential energy of an individual shell, $\rm{d}U$, is then
\begin{eqnarray}
\rm{d} U & = & \frac{-G M(r) \, {\rm d } m }{r} \\
         & = & \frac{-G M(r) \, 4 \pi r^2 \rho(r) \, {\rm d }r}{r} \\
         & = & -16 \pi^2 G \rho_0^2 r_0^4 r
\end{eqnarray}
where $G$ is the gravitational constant.  Integrating this expression
out to a radius $R$ and substituting for $\rho_0$ gives the potential
energy, $U$, for an isothermal sphere
\begin{equation} \label{eq:pe_of_isothermal_sphere}
U = \frac{-G M^2}{R}
\end{equation}

For an ideal gas, the kinetic energy is just
\begin{equation} \label{eq:ke}
K = \frac{3}{2} P V
\end{equation}
where $P$ is the pressure and $V$ is the volume of the
galaxy. Inserting the expressions for the potential and kinetic
energies into the Virial Theorem and dividing by the Boltzmann
constant ($k$) to put the pressure in units of ${\rm K \, cm^{-3}}$,
we obtain the expression
\begin{equation} \label{eq:pressure_unsimplified}
\frac{P}{k} = \frac{G M^2}{3 R V k}.
\end{equation}
Since $V=(4/3) \pi R^3$ and the circular velocity $v$ at radius $R$ in
the galaxy is just $v^2 = GM/R$,
Equation~(\ref{eq:pressure_unsimplified}) can be rewritten as
\begin{equation}
\frac{P}{k} = \frac{v^4 }{4 \pi G R^2 k}.
\end{equation}

We would note that regardless of the potential equation we use
(isothermal sphere, constant density, $\rho \propto r^2$, etc.), the
multiplicative factor in front of the potential changes by only a
factor of 2. Since the gravitational pressure is proportional to the
total potential energy, changes in the assumed form of the potential
only change the gravitational pressure by a factor of a few.

%----------------------------------------------------------------------
%			Bibliography
%----------------------------------------------------------------------

%\bibliography{n4214}

\begin{thebibliography}{88}
\expandafter\ifx\csname natexlab\endcsname\relax\def\natexlab#1{#1}\fi

\bibitem[{{Allsopp}(1979)}]{ch3:1979MNRAS.188..765A}
{Allsopp}, N.~J. 1979, \mnras, 188, 765

\bibitem[{{Beck}(2005)}]{ch3:beck2005}
{Beck}, R. 2005, in Lecture Notes in Physics, Berlin Springer Verlag, Vol. 664,
  Cosmic Magnetic Fields, ed. R.~{Wielebinski} \& R.~{Beck}, 41--+

\bibitem[{{Beck} \& {Gaensler}(2004)}]{2004NewAR..48.1289B}
{Beck}, R., \& {Gaensler}, B.~M. 2004, New Astronomy Review, 48, 1289

\bibitem[{{Beck} {et~al.}(1987){Beck}, {Klein}, \&
  {Wielebinski}}]{1987A&A...186...95B}
{Beck}, R., {Klein}, U., \& {Wielebinski}, R. 1987, \aap, 186, 95

\bibitem[{{Beck} \& {Krause}(2005)}]{ch3:2005AN....326..414B}
{Beck}, R., \& {Krause}, M. 2005, Astronomische Nachrichten, 326, 414

\bibitem[{{Beck} {et~al.}(2003){Beck}, {Shukurov}, {Sokoloff}, \&
  {Wielebinski}}]{ch3:2003A&A...411...99B}
{Beck}, R., {Shukurov}, A., {Sokoloff}, D., \& {Wielebinski}, R. 2003, \aap,
  411, 99

\bibitem[{{Beck} {et~al.}(2000){Beck}, {Turner}, \&
  {Kovo}}]{ch3:2000AJ....120..244B}
{Beck}, S.~C., {Turner}, J.~L., \& {Kovo}, O. 2000, \aj, 120, 244

\bibitem[{{Becker} {et~al.}(1991){Becker}, {White}, \&
  {Edwards}}]{ch3:1991ApJS...75....1B}
{Becker}, R.~H., {White}, R.~L., \& {Edwards}, A.~L. 1991, \apjs, 75, 1

\bibitem[{{Becker} {et~al.}(1995){Becker}, {White}, \&
  {Helfand}}]{1995ApJ...450..559B}
{Becker}, R.~H., {White}, R.~L., \& {Helfand}, D.~J. 1995, \apj, 450, 559

\bibitem[{{Burn}(1966)}]{1966MNRAS.133...67B}
{Burn}, B.~J. 1966, \mnras, 133, 67

\bibitem[{{Chi} \& {Wolfendale}(1993)}]{ch3:cw1993}
{Chi}, X., \& {Wolfendale}, A.~W. 1993, \nat, 362, 610

\bibitem[{{Chyzy} {et~al.}(2000){Chyzy}, {Beck}, {Kohle}, {Klein}, \&
  {Urbanik}}]{ch3:chyzy2000erratum}
{Chyzy}, K.~T., {Beck}, R., {Kohle}, S., {Klein}, U., \& {Urbanik}, M. 2000,
  \aap, 356, 757

\bibitem[{{Chy{\.z}y} {et~al.}(2000){Chy{\.z}y}, {Beck}, {Kohle}, {Klein}, \&
  {Urbanik}}]{ch3:chyzy2000}
{Chy{\.z}y}, K.~T., {Beck}, R., {Kohle}, S., {Klein}, U., \& {Urbanik}, M.
  2000, \aap, 355, 128

\bibitem[{{Chy{\.z}y} {et~al.}(2003){Chy{\.z}y}, {Knapik}, {Bomans}, {Klein},
  {Beck}, {Soida}, \& {Urbanik}}]{ch3:chyzy2003}
{Chy{\.z}y}, K.~T., {Knapik}, J., {Bomans}, D.~J., {Klein}, U., {Beck}, R.,
  {Soida}, M., \& {Urbanik}, M. 2003, \aap, 405, 513

\bibitem[{{Chy{\.z}y} {et~al.}(2011){Chy{\.z}y}, {We{\.z}gowiec}, {Beck}, \&
  {Bomans}}]{2011A&A...529A..94C}
{Chy{\.z}y}, K.~T., {We{\.z}gowiec}, M., {Beck}, R., \& {Bomans}, D.~J. 2011,
  \aap, 529, A94+

\bibitem[{{Condon} {et~al.}(2002){Condon}, {Cotton}, \&
  {Broderick}}]{ch3:2002AJ....124..675C}
{Condon}, J.~J., {Cotton}, W.~D., \& {Broderick}, J.~J. 2002, \aj, 124, 675

\bibitem[{{Cormier} {et~al.}(2010){Cormier}, {Madden}, {Hony}, {Contursi},
  {Poglitsch}, {Galliano}, {Sturm}, {Doublier}, {Feuchtgruber}, {Galametz},
  {Geis}, {de Jong}, {Okumura}, {Panuzzo}, \& {Sauvage}}]{2010A&A...518L..57C}
{Cormier}, D., {Madden}, S.~C., {Hony}, S., {Contursi}, A., {Poglitsch}, A.,
  {Galliano}, F., {Sturm}, E., {Doublier}, V., {Feuchtgruber}, H., {Galametz},
  M., {Geis}, N., {de Jong}, J., {Okumura}, K., {Panuzzo}, P., \& {Sauvage}, M.
  2010, \aap, 518, L57+

\bibitem[{{Cox}(2005)}]{2005ARA&A..43..337C}
{Cox}, D.~P. 2005, \araa, 43, 337

\bibitem[{{Dalcanton} {et~al.}(2009){Dalcanton}, {Williams}, {Seth}, {Dolphin},
  {Holtzman}, {Rosema}, {Skillman}, {Cole}, {Girardi}, {Gogarten},
  {Karachentsev}, {Olsen}, {Weisz}, {Christensen}, {Freeman}, {Gilbert},
  {Gallart}, {Harris}, {Hodge}, {de Jong}, {Karachentseva}, {Mateo}, {Stetson},
  {Tavarez}, {Zaritsky}, {Governato}, \& {Quinn}}]{2009ApJS..183...67D}
{Dalcanton}, J.~J., {Williams}, B.~F., {Seth}, A.~C., {Dolphin}, A.,
  {Holtzman}, J., {Rosema}, K., {Skillman}, E.~D., {Cole}, A., {Girardi}, L.,
  {Gogarten}, S.~M., {Karachentsev}, I.~D., {Olsen}, K., {Weisz}, D.,
  {Christensen}, C., {Freeman}, K., {Gilbert}, K., {Gallart}, C., {Harris}, J.,
  {Hodge}, P., {de Jong}, R.~S., {Karachentseva}, V., {Mateo}, M., {Stetson},
  P.~B., {Tavarez}, M., {Zaritsky}, D., {Governato}, F., \& {Quinn}, T. 2009,
  \apjs, 183, 67

\bibitem[{{de Avillez} \& {Breitschwerdt}(2005)}]{ch2:2005A&A...436..585D}
{de Avillez}, M.~A., \& {Breitschwerdt}, D. 2005, \aap, 436, 585

\bibitem[{{de Vaucouleurs} {et~al.}(1995){de Vaucouleurs}, {de Vaucouleurs},
  {Corwin}, {Buta}, {Paturel}, \& {Fouque}}]{ch3:1995yCat.7155....0D}
{de Vaucouleurs}, G., {de Vaucouleurs}, A., {Corwin}, H.~G., {Buta}, R.~J.,
  {Paturel}, G., \& {Fouque}, P. 1995, VizieR Online Data Catalog, 7155, 0

\bibitem[{{Deeg} {et~al.}(1993){Deeg}, {Brinks}, {Duric}, {Klein}, \&
  {Skillman}}]{1993ApJ...410..626D}
{Deeg}, H.-J., {Brinks}, E., {Duric}, N., {Klein}, U., \& {Skillman}, E. 1993,
  \apj, 410, 626

\bibitem[{{Dogiel} \& {Breitschwerdt}(2009)}]{2009arXiv0905.3071D}
{Dogiel}, V., \& {Breitschwerdt}, D. 2009, ArXiv e-prints

\bibitem[{{Dopita} {et~al.}(2010){Dopita}, {Calzetti}, {Ma{\'{\i}}z
  Apell{\'a}niz}, {Blair}, {Long}, {Mutchler}, {Whitmore}, {Bond}, {MacKenty},
  {Balick}, {Carollo}, {Disney}, {Frogel}, {O'Connell}, {Hall}, {Holtzman},
  {Kimble}, {McCarthy}, {Paresce}, {Saha}, {Walker}, {Silk}, {Sirianni},
  {Trauger}, {Windhorst}, \& {Young}}]{2010Ap&SS.tmp..148D}
{Dopita}, M.~A., {Calzetti}, D., {Ma{\'{\i}}z Apell{\'a}niz}, J., {Blair},
  W.~P., {Long}, K.~S., {Mutchler}, M., {Whitmore}, B.~C., {Bond}, H.~E.,
  {MacKenty}, J., {Balick}, B., {Carollo}, M., {Disney}, M., {Frogel}, J.~A.,
  {O'Connell}, R., {Hall}, D., {Holtzman}, J.~A., {Kimble}, R.~A., {McCarthy},
  P., {Paresce}, F., {Saha}, A., {Walker}, A.~R., {Silk}, J., {Sirianni}, M.,
  {Trauger}, J., {Windhorst}, R., \& {Young}, E. 2010, \apss, 148

\bibitem[{{Dressel} \& {Condon}(1978)}]{ch3:1978ApJS...36...53D}
{Dressel}, L.~L., \& {Condon}, J.~J. 1978, \apjs, 36, 53

\bibitem[{{Drozdovsky} {et~al.}(2002){Drozdovsky}, {Schulte-Ladbeck}, {Hopp},
  {Greggio}, \& {Crone}}]{ch3:2002AJ....124..811D}
{Drozdovsky}, I.~O., {Schulte-Ladbeck}, R.~E., {Hopp}, U., {Greggio}, L., \&
  {Crone}, M.~M. 2002, \aj, 124, 811

\bibitem[{{Everett} {et~al.}(2010){Everett}, {Schiller}, \&
  {Zweibel}}]{2010ApJ...711...13E}
{Everett}, J.~E., {Schiller}, Q.~G., \& {Zweibel}, E.~G. 2010, \apj, 711, 13

\bibitem[{{Everett} {et~al.}(2008){Everett}, {Zweibel}, {Benjamin}, {McCammon},
  {Rocks}, \& {Gallagher}}]{2008ApJ...674..258E}
{Everett}, J.~E., {Zweibel}, E.~G., {Benjamin}, R.~A., {McCammon}, D., {Rocks},
  L., \& {Gallagher}, III, J.~S. 2008, \apj, 674, 258

\bibitem[{{Fanelli} {et~al.}(1997){Fanelli}, {Waller}, {Smith}, {Freedman},
  {Madore}, {Neff}, {O'Connell}, {Roberts}, {Bohlin}, {Smith}, \&
  {Stecher}}]{ch3:1997ApJ...481..735F}
{Fanelli}, M.~N., {Waller}, W.~W., {Smith}, D.~A., {Freedman}, W.~L., {Madore},
  B., {Neff}, S.~G., {O'Connell}, R.~W., {Roberts}, M.~S., {Bohlin}, R.,
  {Smith}, A.~M., \& {Stecher}, T.~P. 1997, \apj, 481, 735

\bibitem[{{Ferrara} \& {Tolstoy}(2000)}]{2000MNRAS.313..291F}
{Ferrara}, A., \& {Tolstoy}, E. 2000, \mnras, 313, 291

\bibitem[{{Ferri{\`e}re}(2001)}]{2001RvMP...73.1031F}
{Ferri{\`e}re}, K.~M. 2001, Reviews of Modern Physics, 73, 1031

\bibitem[{{Gaensler} {et~al.}(2005){Gaensler}, {Haverkorn}, {Staveley-Smith},
  {Dickey}, {McClure-Griffiths}, {Dickel}, \& {Wolleben}}]{ch3:gaensler2005}
{Gaensler}, B.~M., {Haverkorn}, M., {Staveley-Smith}, L., {Dickey}, J.~M.,
  {McClure-Griffiths}, N.~M., {Dickel}, J.~R., \& {Wolleben}, M. 2005, Science,
  307, 1610

\bibitem[{{Gregory} \& {Condon}(1991)}]{ch3:1991ApJS...75.1011G}
{Gregory}, P.~C., \& {Condon}, J.~J. 1991, \apjs, 75, 1011

\bibitem[{{Greisen}(2008)}]{ch3:aipscookbook}
{Greisen}, E., ed. 2008, {AIPS Cookbook} (NRAO)

\bibitem[{{Hanasz} {et~al.}(2004){Hanasz}, {Kowal}, {Otmianowska-Mazur}, \&
  {Lesch}}]{ch3:2004ApJ...605L..33H}
{Hanasz}, M., {Kowal}, G., {Otmianowska-Mazur}, K., \& {Lesch}, H. 2004, \apjl,
  605, L33

\bibitem[{{Hartwell} {et~al.}(2004){Hartwell}, {Stevens}, {Strickland},
  {Heckman}, \& {Summers}}]{ch3:2004MNRAS.348..406H}
{Hartwell}, J.~M., {Stevens}, I.~R., {Strickland}, D.~K., {Heckman}, T.~M., \&
  {Summers}, L.~K. 2004, \mnras, 348, 406

\bibitem[{{Haynes} {et~al.}(1991){Haynes}, {Klein}, {Wayte}, {Wielebinski},
  {Murray}, {Bajaja}, {Meinert}, {Buczilowski}, {Harnett}, {Hunt}, {Wark}, \&
  {Sciacca}}]{ch3:1991A&A...252..475H}
{Haynes}, R.~F., {Klein}, U., {Wayte}, S.~R., {Wielebinski}, R., {Murray},
  J.~D., {Bajaja}, E., {Meinert}, D., {Buczilowski}, U.~R., {Harnett}, J.~I.,
  {Hunt}, A.~J., {Wark}, R., \& {Sciacca}, L. 1991, \aap, 252, 475

\bibitem[{{Haynes} {et~al.}(1986){Haynes}, {Murray}, {Klein}, \&
  {Wielebinski}}]{ch3:1986A&A...159...22H}
{Haynes}, R.~F., {Murray}, J.~D., {Klein}, U., \& {Wielebinski}, R. 1986, \aap,
  159, 22

\bibitem[{{Hopp} {et~al.}(1999){Hopp}, {Schulte-Ladbeck}, {Greggio}, \&
  {Crone}}]{ch3:1999ASPC..192...85H}
{Hopp}, U., {Schulte-Ladbeck}, R.~E., {Greggio}, L., \& {Crone}, M.~M. 1999, in
  Astronomical Society of the Pacific Conference Series, Vol. 192,
  Spectrophotometric Dating of Stars and Galaxies, ed. I.~{Hubeny}, S.~{Heap},
  \& R.~{Cornett}, 85--+

\bibitem[{{Hunt} {et~al.}(2004){Hunt}, {Dyer}, {Thuan}, \&
  {Ulvestad}}]{ch3:2004ApJ...606..853H}
{Hunt}, L.~K., {Dyer}, K.~K., {Thuan}, T.~X., \& {Ulvestad}, J.~S. 2004, \apj,
  606, 853

\bibitem[{{Hunter} {et~al.}(2007){Hunter}, {Brinks}, {Elmegreen}, {Rupen},
  {Simpson}, {Walter}, {Westpfahl}, \& {Young}}]{ch3:2007AAS...211.9506H}
{Hunter}, D.~A., {Brinks}, E., {Elmegreen}, B., {Rupen}, M., {Simpson}, C.,
  {Walter}, F., {Westpfahl}, D., \& {Young}, L. 2007, in American Astronomical
  Society Meeting Abstracts, Vol. 211, American Astronomical Society Meeting
  Abstracts, \#95.06

\bibitem[{{Hunter} \& {Elmegreen}(2004)}]{ch3:hunter2004}
{Hunter}, D.~A., \& {Elmegreen}, B.~G. 2004, \aj, 128, 2170

\bibitem[{{Hunter} \& {Elmegreen}(2006)}]{ch3:hunter2006ubv}
---. 2006, \apjs, 162, 49

\bibitem[{{Hunter} {et~al.}(2010){Hunter}, {Elmegreen}, \&
  {Ludka}}]{2010AJ....139..447H}
{Hunter}, D.~A., {Elmegreen}, B.~G., \& {Ludka}, B.~C. 2010, \aj, 139, 447

\bibitem[{{Hunter} {et~al.}(2006){Hunter}, {Elmegreen}, \&
  {Martin}}]{ch3:hunter2006ir}
{Hunter}, D.~A., {Elmegreen}, B.~G., \& {Martin}, E. 2006, \aj, 132, 801

\bibitem[{{Kepley} {et~al.}(2010){Kepley}, {M{\"u}hle}, {Everett}, {Zweibel},
  {Wilcots}, \& {Klein}}]{2010ApJ...712..536K}
{Kepley}, A.~A., {M{\"u}hle}, S., {Everett}, J., {Zweibel}, E.~G., {Wilcots},
  E.~M., \& {Klein}, U. 2010, \apj, 712, 536

\bibitem[{{Kepley} {et~al.}(2007){Kepley}, {Wilcots}, {Hunter}, \&
  {Nordgren}}]{2007AJ....133.2242K}
{Kepley}, A.~A., {Wilcots}, E.~M., {Hunter}, D.~A., \& {Nordgren}, T. 2007,
  \aj, 133, 2242

\bibitem[{{Klein} {et~al.}(1993){Klein}, {Haynes}, {Wielebinski}, \&
  {Meinert}}]{ch3:1993A&A...271..402K}
{Klein}, U., {Haynes}, R.~F., {Wielebinski}, R., \& {Meinert}, D. 1993, \aap,
  271, 402

\bibitem[{{Klein} {et~al.}(1984){Klein}, {Wielebinski}, \&
  {Thuan}}]{1984A&A...141..241K}
{Klein}, U., {Wielebinski}, R., \& {Thuan}, T.~X. 1984, \aap, 141, 241

\bibitem[{{Kobulnicky} \& {Skillman}(1996)}]{ch3:1996ApJ...471..211K}
{Kobulnicky}, H.~A., \& {Skillman}, E.~D. 1996, \apj, 471, 211

\bibitem[{{Kulsrud}(1999)}]{1999ARA&A..37...37K}
{Kulsrud}, R.~M. 1999, \araa, 37, 37

\bibitem[{{Kulsrud} \& {Zweibel}(2008)}]{2008RPPh...71d6901K}
{Kulsrud}, R.~M., \& {Zweibel}, E.~G. 2008, Reports on Progress in Physics, 71,
  046901

\bibitem[{{Lisenfeld} {et~al.}(2008){Lisenfeld}, {Relano}, {Vilchez},
  {Battaner}, \& {Hermelo}}]{2008arXiv0808.1953L}
{Lisenfeld}, U., {Relano}, M., {Vilchez}, J., {Battaner}, E., \& {Hermelo}, I.
  2008, ArXiv e-prints, 808

\bibitem[{{MacKenty} {et~al.}(2000){MacKenty}, {Ma{\'{\i}}z-Apell{\'a}niz},
  {Pickens}, {Norman}, \& {Walborn}}]{ch3:2000AJ....120.3007M}
{MacKenty}, J.~W., {Ma{\'{\i}}z-Apell{\'a}niz}, J., {Pickens}, C.~E., {Norman},
  C.~A., \& {Walborn}, N.~R. 2000, \aj, 120, 3007

\bibitem[{{Ma{\'{\i}}z-Apell{\'a}niz}
  {et~al.}(2002){Ma{\'{\i}}z-Apell{\'a}niz}, {Cieza}, \&
  {MacKenty}}]{ch3:2002AJ....123.1307M}
{Ma{\'{\i}}z-Apell{\'a}niz}, J., {Cieza}, L., \& {MacKenty}, J.~W. 2002, \aj,
  123, 1307

\bibitem[{{Maiz-Apellaniz} {et~al.}(1998){Maiz-Apellaniz}, {Mas-Hesse},
  {Munoz-Tunon}, {Vilchez}, \& {Castaneda}}]{ch3:1998A&A...329..409M}
{Maiz-Apellaniz}, J., {Mas-Hesse}, J.~M., {Munoz-Tunon}, C., {Vilchez}, J.~M.,
  \& {Castaneda}, H.~O. 1998, \aap, 329, 409

\bibitem[{{Ma{\'{\i}}z-Apell{\'a}niz}
  {et~al.}(1999){Ma{\'{\i}}z-Apell{\'a}niz}, {Mu{\~n}oz-Tu{\~n}{\'o}n},
  {Tenorio-Tagle}, \& {Mas-Hesse}}]{ch3:1999A&A...343...64M}
{Ma{\'{\i}}z-Apell{\'a}niz}, J., {Mu{\~n}oz-Tu{\~n}{\'o}n}, C.,
  {Tenorio-Tagle}, G., \& {Mas-Hesse}, J.~M. 1999, \aap, 343, 64

\bibitem[{{Mao} {et~al.}(2008){Mao}, {Gaensler}, {Stanimirovi{\'c}},
  {Haverkorn}, {McClure-Griffiths}, {Staveley-Smith}, \&
  {Dickey}}]{ch3:mao2008}
{Mao}, S.~A., {Gaensler}, B.~M., {Stanimirovi{\'c}}, S., {Haverkorn}, M.,
  {McClure-Griffiths}, N.~M., {Staveley-Smith}, L., \& {Dickey}, J.~M. 2008,
  \apj, accepted

\bibitem[{{Marasco} \& {Fraternali}(2011)}]{2011A&A...525A.134M}
{Marasco}, A., \& {Fraternali}, F. 2011, \aap, 525, A134+

\bibitem[{{Martin}(1998)}]{ch3:martin98}
{Martin}, C.~L. 1998, \apj, 506, 222

\bibitem[{{McIntyre}(1997)}]{ch3:mcintyre97}
{McIntyre}, V.~J. 1997, Publications of the Astronomical Society of Australia,
  14, 122

\bibitem[{{McMahon} {et~al.}(2002){McMahon}, {White}, {Helfand}, \&
  {Becker}}]{2002ApJS..143....1M}
{McMahon}, R.~G., {White}, R.~L., {Helfand}, D.~J., \& {Becker}, R.~H. 2002,
  \apjs, 143, 1

\bibitem[{{Mineshige} {et~al.}(1993){Mineshige}, {Shibata}, \&
  {Shapiro}}]{ch2:1993ApJ...409..663M}
{Mineshige}, S., {Shibata}, K., \& {Shapiro}, P.~R. 1993, \apj, 409, 663

\bibitem[{{Otmianowska-Mazur} {et~al.}(2000){Otmianowska-Mazur}, {Chy{\.z}y},
  {Soida}, \& {von Linden}}]{ch3:om2000}
{Otmianowska-Mazur}, K., {Chy{\.z}y}, K.~T., {Soida}, M., \& {von Linden}, S.
  2000, \aap, 359, 29

\bibitem[{{Ott} {et~al.}(2005{\natexlab{a}}){Ott}, {Walter}, \&
  {Brinks}}]{ch3:2005MNRAS.358.1423O}
{Ott}, J., {Walter}, F., \& {Brinks}, E. 2005{\natexlab{a}}, \mnras, 358, 1423

\bibitem[{{Ott} {et~al.}(2005{\natexlab{b}}){Ott}, {Walter}, \&
  {Brinks}}]{ch3:ott05II}
---. 2005{\natexlab{b}}, \mnras, 358, 1453

\bibitem[{{Ott} {et~al.}(2001){Ott}, {Walter}, {Brinks}, {Van Dyk}, {Dirsch},
  \& {Klein}}]{ch3:2001AJ....122.3070O}
{Ott}, J., {Walter}, F., {Brinks}, E., {Van Dyk}, S.~D., {Dirsch}, B., \&
  {Klein}, U. 2001, \aj, 122, 3070

\bibitem[{{Pacholczyk}(1970)}]{ch3:1970ranp.book.....P}
{Pacholczyk}, A.~G. 1970, {Radio astrophysics. Nonthermal processes in galactic
  and extragalactic sources} (Series of Books in Astronomy and Astrophysics,
  San Francisco: Freeman, 1970)

\bibitem[{{Parker}(1992)}]{ch3:1992ApJ...401..137P}
{Parker}, E.~N. 1992, \apj, 401, 137

\bibitem[{{Sault} {et~al.}(1995){Sault}, {Teuben}, \&
  {Wright}}]{ch3:1995ASPC...77..433S}
{Sault}, R.~J., {Teuben}, P.~J., \& {Wright}, M.~C.~H. 1995, in Astronomical
  Society of the Pacific Conference Series, Vol.~77, Astronomical Data Analysis
  Software and Systems IV, ed. R.~A. {Shaw}, H.~E. {Payne}, \& J.~J.~E.
  {Hayes}, 433--+

\bibitem[{{Schlegel} {et~al.}(1998){Schlegel}, {Finkbeiner}, \&
  {Davis}}]{ch3:1998ApJ...500..525S}
{Schlegel}, D.~J., {Finkbeiner}, D.~P., \& {Davis}, M. 1998, \apj, 500, 525

\bibitem[{{Schmitt} {et~al.}(2006){Schmitt}, {Calzetti}, {Armus}, {Giavalisco},
  {Heckman}, {Kennicutt}, {Leitherer}, \& {Meurer}}]{ch3:2006ApJS..164...52S}
{Schmitt}, H.~R., {Calzetti}, D., {Armus}, L., {Giavalisco}, M., {Heckman},
  T.~M., {Kennicutt}, Jr., R.~C., {Leitherer}, C., \& {Meurer}, G.~R. 2006,
  \apjs, 164, 52

\bibitem[{{Schwartz} \& {Martin}(2004)}]{ch3:schwartz04}
{Schwartz}, C.~M., \& {Martin}, C.~L. 2004, \apj, 610, 201

\bibitem[{{Simmons} \& {Stewart}(1985)}]{ch2:simmons1985}
{Simmons}, J.~F.~L., \& {Stewart}, B.~G. 1985, \aap, 142, 100

\bibitem[{{Sokoloff} {et~al.}(1998){Sokoloff}, {Bykov}, {Shukurov},
  {Berkhuijsen}, {Beck}, \& {Poezd}}]{ch3:1998MNRAS.299..189S}
{Sokoloff}, D.~D., {Bykov}, A.~A., {Shukurov}, A., {Berkhuijsen}, E.~M.,
  {Beck}, R., \& {Poezd}, A.~D. 1998, \mnras, 299, 189

\bibitem[{{Spitzer}(1978)}]{1978ppim.book.....S}
{Spitzer}, L. 1978, {Physical processes in the interstellar medium}, ed.
  {Spitzer, L.}

\bibitem[{{Springel} {et~al.}(2005){Springel}, {White}, {Jenkins}, {Frenk},
  {Yoshida}, {Gao}, {Navarro}, {Thacker}, {Croton}, {Helly}, {Peacock}, {Cole},
  {Thomas}, {Couchman}, {Evrard}, {Colberg}, \&
  {Pearce}}]{ch3:2005Natur.435..629S}
{Springel}, V., {White}, S.~D.~M., {Jenkins}, A., {Frenk}, C.~S., {Yoshida},
  N., {Gao}, L., {Navarro}, J., {Thacker}, R., {Croton}, D., {Helly}, J.,
  {Peacock}, J.~A., {Cole}, S., {Thomas}, P., {Couchman}, H., {Evrard}, A.,
  {Colberg}, J., \& {Pearce}, F. 2005, \nat, 435, 629

\bibitem[{{Steidel} {et~al.}(1996){Steidel}, {Giavalisco}, {Pettini},
  {Dickinson}, \& {Adelberger}}]{ch3:1996ApJ...462L..17S}
{Steidel}, C.~C., {Giavalisco}, M., {Pettini}, M., {Dickinson}, M., \&
  {Adelberger}, K.~L. 1996, \apjl, 462, L17+

\bibitem[{{Subramanian}(1998)}]{ch3:1998MNRAS.294..718S}
{Subramanian}, K. 1998, \mnras, 294, 718

\bibitem[{{Tomisaka}(1990)}]{ch2:1990ApJ...361L...5T}
{Tomisaka}, K. 1990, \apjl, 361, L5

\bibitem[{{Tremonti} {et~al.}(2004){Tremonti}, {Heckman}, {Kauffmann},
  {Brinchmann}, {Charlot}, {White}, {Seibert}, {Peng}, {Schlegel}, {Uomoto},
  {Fukugita}, \& {Brinkmann}}]{2004ApJ...613..898T}
{Tremonti}, C.~A., {Heckman}, T.~M., {Kauffmann}, G., {Brinchmann}, J.,
  {Charlot}, S., {White}, S.~D.~M., {Seibert}, M., {Peng}, E.~W., {Schlegel},
  D.~J., {Uomoto}, A., {Fukugita}, M., \& {Brinkmann}, J. 2004, \apj, 613, 898

\bibitem[{{{\'U}beda} {et~al.}(2007{\natexlab{a}}){{\'U}beda},
  {Ma{\'{\i}}z-Apell{\'a}niz}, \& {MacKenty}}]{ch3:2007AJ....133..917U}
{{\'U}beda}, L., {Ma{\'{\i}}z-Apell{\'a}niz}, J., \& {MacKenty}, J.~W.
  2007{\natexlab{a}}, \aj, 133, 917

\bibitem[{{{\'U}beda} {et~al.}(2007{\natexlab{b}}){{\'U}beda},
  {Ma{\'{\i}}z-Apell{\'a}niz}, \& {MacKenty}}]{ch3:2007AJ....133..932U}
---. 2007{\natexlab{b}}, \aj, 133, 932

\bibitem[{{Vaillancourt}(2006)}]{ch2:2006PASP..118.1340V}
{Vaillancourt}, J.~E. 2006, \pasp, 118, 1340

\bibitem[{{van Eymeren} {et~al.}(2009){van Eymeren}, {Trachternach},
  {Koribalski}, \& {Dettmar}}]{2009A&A...505....1V}
{van Eymeren}, J., {Trachternach}, C., {Koribalski}, B.~S., \& {Dettmar}, R.
  2009, \aap, 505, 1

\bibitem[{{Wielebinski}(2005)}]{2005LNP...664...89W}
{Wielebinski}, R. 2005, in Lecture Notes in Physics, Berlin Springer Verlag,
  Vol. 664, Cosmic Magnetic Fields, ed. {R.~Wielebinski \& R.~Beck}, 89--+

\bibitem[{{Wilcots} \& {Thurow}(2001)}]{ch3:2001ApJ...555..758W}
{Wilcots}, E.~M., \& {Thurow}, J.~C. 2001, \apj, 555, 758

\bibitem[{{Zweibel} \& {Heiles}(1997)}]{ch3:zh97}
{Zweibel}, E.~G., \& {Heiles}, C. 1997, \nat, 385, 131

\end{thebibliography}

%----------------------------------------------------------------------
%			        Tables
%----------------------------------------------------------------------

\clearpage

%%% Table 1
%\input{tables/vla_obs_summary.tex}
\begin{deluxetable}{llll}
\tablewidth{0pt}
\tablecaption{Summary of VLA Observations \label{tab:vla_obs_summary}}
\tablehead{ \colhead{Observing Band} &
\colhead{20cm } &
\colhead{6cm } &
\colhead{3cm } }
\startdata
Array						&     C		 &      D		 &	 D                    \\
Date						& 2005 Aug 7	 & 2005 Dec 5 \& 19	 &  2005 Dec 23 \& 24     \\
IF 1 Frequency (GHz)				& 1.3649	 & 4.8851		 & 8.4351                     \\
IF 2 Frequency (GHz)				& 1.4351	 & 4.8351		 & 8.4851                     \\
Bandwidth/IF (MHz)					& 50		 & 50			 & 50                         \\ 
Number of Channels/IF				& 1		 & 1			 & 1                          \\
Primary Calibrator				& 1331+305	 & 1331+305		 & 1331+305                   \\
Secondary Calibrator				& 1227+365	 & 1146+399		 & 1146+399                   \\
Number of Pointings				&  1		 & 4			 & 16                         \\
Field of View (\arcmin)                         &  30            & 9                     & 5.4                          \\ 
Resolution (\arcsec)			& 12.5		 & 14.0			 & 8.4                        \\ 
Largest Angular Scale (\arcmin)			& 15		 & 5			 & 3                          \\
Integration Time per Pointing (hours)		& 10.0		 & 3.73			 & 1.2                        
\enddata
\end{deluxetable}

%%% Table 2
%%\input{tables/final_image_summary.tex}
\begin{deluxetable}{rccccc}
\tablewidth{0pt}
\tabletypesize{\scriptsize}
\tablecaption{Final Images \label{tab:final_image_summary}} 
\tablehead{ 
  \colhead{Frequency} &
  \colhead{Beam} &
  \colhead{PA} &
  \colhead{$\sigma_I$}  &
  \colhead{$\sigma_Q$} &
  \colhead{$\sigma_U$} \\
  \colhead{GHz} &
  \colhead{$\arcsec$} &
  \colhead{$^\circ$} &
  \colhead{$\mu$Jy beam$^{-1}$} &
  \colhead{$\mu$Jy beam$^{-1}$} &
  \colhead{$\mu$Jy beam$^{-1}$} 
} 
\startdata
1.397      & $14.18\arcsec \times 11.46 \arcsec$ & $-77.3^\circ$ & 48.8 & 13.0 & 12.9  \\
4.860      & $14.18\arcsec \times 11.46 \arcsec$ & $-77.3^\circ$ & 25.1 & 17.8 & 18.5  \\
8.460      & $14.18\arcsec \times 11.46 \arcsec$ & $-77.3^\circ$ & 24.1 & 20.7 & 21.0  \\
\enddata
\end{deluxetable}

%%% Table 3
%%\input{tables/radio_cont_summary.tex}
\begin{deluxetable}{ccccc}
\tablewidth{0pt}
\tablecaption{Radio Continuum Flux Spectrum \label{tab:radio_cont_spectrum}}
\tablehead{\colhead{Frequency} &
  \colhead{Flux} &
  \colhead{Flux Error} &
  \colhead{} &
  \colhead{} \\
  \colhead{GHz} &
  \colhead{mJy} &
  \colhead{mJy} &
  \colhead{Telescope} & 
  \colhead{Source} }
\startdata
%% Freq (GHZ)    Flux (mJy)    Flux Error (mJy)    Telescope        Source 
1.40	&	  51.5	   &	   10.3          &  VLA        &        6      \\
1.40	&	  38.3	   &	   7.7\phn	 &  VLA/NVSS   &	5      \\
2.38	&	  36.0     &       3.0\phn       &  Arecibo    &        4      \\
4.86	&         34.0	   &	   6.8	         &  VLA        &	6      \\
4.85	&	  30.0     &       4.5\phn       &  GB 300ft   &        2      \\
4.85    &         30.0     &       7.0\phn	 &  GB 300ft   &        3      \\
8.46	&	  20.5	   &	   0.5\phn       &  VLA        &        1      \\
8.46	&         24.2	   &	   4.8	         &  VLA        &        6      \\
\enddata
\tablerefs{(1) \citealt{ch3:2006ApJS..164...52S}; (2) \citealt{ch3:1991ApJS...75....1B}; (3) \citealt{ch3:1991ApJS...75.1011G}; (4) \citealt{ch3:1978ApJS...36...53D}; (5) \citealt{ch3:2002AJ....124..675C}; (6) This paper  }
\end{deluxetable}

%%% Table 4
%%\input{tables/ism_pressures.tex}
\begin{deluxetable}{lcccc}
\tablewidth{0pt}
\tabletypesize{\scriptsize}
\tablecaption{Pressures of Various Components of the ISM of NGC 4214 \label{tab:ism_pressures}}
\tablehead{
\colhead{} &
\colhead{} &
\colhead{} &
\colhead{} &
\colhead{Pressure} \\ 
\colhead{Component} &
\colhead{Input Values} & 
\colhead{Equation} &
\colhead{Reference} & 
\colhead{$10^5\ \rm{K} \ \rm{cm^{-3}}$} }
\startdata
Magnetic field (center)  &   $B = 30 \, \uG$                                                   & $B^2/(8 \pi k)$                               & 1 & 2.6   \\
Magnetic field (edges)   &   $B = 10 \, \uG$                                                   & $B^2/(8 \pi k)$                               & 1 & 0.3   \\
Magnetic field (total)   &   $B = 13 \, \uG$                                                   & $B^2/(8 \pi k)$                               & 1 & 0.5   \\
Hot gas (central)        &   $n_e = 0.147 \, {\rm cm^{-3}}, \ T= 2.80 \times 10^6 \, {\rm K}$  & $2 n_e T$                                     & 2 & 8.2   \\
Hot gas (edges)          &   $n_e = 0.033 \, {\rm cm^{-3}}, \ T= 2.68 \times 10^6 \, {\rm K}$  & $2 n_e T$                                     & 2 & 1.7   \\
Hot gas (total)          &   $n_e = 0.064 \, {\rm cm^{-3}}, \ T= 2.17 \times 10^6 \, {\rm K}$  & $2 n_e T$                                     & 2 & 2.8   \\
HII regions (thermal)    &   $n_e = 100 \, {\rm cm^{-3}}, \ T = 10,500 \, {\rm K}$             & $2 n_e T$                                     & 3 & 10.5  \\
HII regions (turbulent)  &   $n_e = 100 \, {\rm cm^{-3}}, \ v_{turb} = 50 \, \kms$             & $n_e m_H v_{turb}^2 / ( k)$                   & 4 & 300   \\
Gravity\tablenotemark{a} &   $v_{rad} = 20\, \kms, i = 30\arcdeg, \ r= 1.11\, {\rm kpc}$       & $(v_{rad}/\sin{i})^4 / (4 \pi r^2 G k)$      & 5 &  1.9 \\
\enddata
\tablenotetext{a}{See Appendix~\ref{sec:deriv-grav-press} for a derivation of the expression used to calculate the graviational pressure.}
\tablerefs{ (1) This work; (2) \citet{ch3:ott05II}; (3) \citet{ch3:1996ApJ...471..211K}; (4) \citet{ch3:2001ApJ...555..758W}; (5) \citet{ch3:1979MNRAS.188..765A}. }
\end{deluxetable}

%----------------------------------------------------------------------
%			        Figures
%----------------------------------------------------------------------

\clearpage

%%% Figure 1

%%% Figure 2

%%% Figure 3

%%% Figure 4

%%% Figure 5

%%% Figure 6

%%% Figure 7

%%% Figure 8

%%% Figure 9

%%% Figure 10

%%% Figure 11

\end{document}